\documentclass[journal]{IEEEtran}

\usepackage{amsmath, amsfonts, amsthm}
\usepackage{hyperref}
\usepackage{physics}
\usepackage{bm}
\usepackage{soul, color, xcolor}
\usepackage{amsfonts}
\usepackage{array}
\usepackage{indentfirst}
\usepackage{algorithmic}
\usepackage{algorithm}
\usepackage{enumerate}
\usepackage{array}
\usepackage[caption=false, font=normalsize, labelfont=sf, textfont=sf]{subfig}
\usepackage{textcomp}
\usepackage{stfloats}
\usepackage{caption}
\usepackage{booktabs}
\usepackage{multirow}
\usepackage{tabularx}
\usepackage{url}
\usepackage{float}
\usepackage{verbatim}
\usepackage{graphicx}
\usepackage{cite}
\usepackage{cuted}
\usepackage{xspace}
\usepackage{url}

\newtheorem{theorem}{Theorem}

\newcommand{\Nir}[1]{\textcolor{red}{\footnotesize{\textsf {[Nir: #1]}}}}

\newcommand{\nameeot}{EOTNet\xspace}

\usepackage[all=normal,paragraphs=tight,floats=tight,mathspacing=normal,wordspacing=tight,charwidths=tight,mathdisplays=normal,leading=normal]{savetrees}

\theoremstyle{remark}

\begin{document}

\title{\nameeot: Deep Memory Aided Bayesian Filter\\ for Extended Object Tracking}
\author{Zhixing Wang, Le Zheng,~\IEEEmembership{Senior Member,~IEEE,} Shi Yan, Ruud J. G. van Sloun,~\IEEEmembership{Member,~IEEE,} \\Nir Shlezinger,~\IEEEmembership{Senior Member,~IEEE,} and Yonina C. Eldar,~\IEEEmembership{Fellow,~IEEE} \vspace{-2em} 
\thanks{Zhixing Wang and Le Zheng are with the Radar Research Laboratory, School of Information and Electronics, Beijing Institute of Technology, Beijing 100081, China (e-mail: wangwangzx@bit.edu.cn; le.zheng.cn@gmail.com).}
\thanks{Shi Yan is with the School of Automation, Northwestern Polytechnical University, Xi’an 710072, China (e-mail: yanshi@mail.nwpu.edu.cn).}
\thanks{Ruud J. G. van Sloun is with the Department of Electrical Engineering, Eindhoven University of Technology, 5612 Eindhoven, The Netherlands (e-mail: r.j.g.v.sloun@tue.nl).}
\thanks{Nir Shlezinger is with the School of Electrical and Computer Engineering, Ben-Gurion University of the Negev, Beer Sheva 84105, Israel (e-mail: nirshl@bgu.ac.il).}
\thanks{Yonina C. Eldar is with the Faculty of Math and CS, Weizmann Institute of Science, Rehovot 7610001, Israel (e-mail: yonina.eldar@weizmann.ac.il).}

} 

\markboth{Journal of \LaTeX\ Class Files, ~Vol.~14, No.~8, August~2021}
{Shell \MakeLowercase{\textit{et al.}}: A Sample Article Using IEEEtran.cls for IEEE Journals}


\maketitle
\begin{abstract}
Extended object tracking methods based on random matrices, founded on Bayesian filters, have been able to achieve efficient recursive processes while jointly estimating the kinematic states and extension of the targets. Existing random matrix approaches typically assume that the evolution of state and extension follows a first-order Markov process, where the current estimate of the target depends solely on the previous moment. However, in real-world scenarios, this assumption fails because the evolution of states and extension is usually non-Markovian. In this paper, we introduce a novel extended object tracking method: a Bayesian recursive neural network assisted by deep memory. Initially, we propose an equivalent model under a non-Markovian assumption and derive the implementation of its Bayesian filtering framework. Thereafter, Gaussian approximation and moment matching are employed to derive the analytical solution for the proposed Bayesian filtering framework. Finally, based on the closed-form solution, we design an end-to-end trainable Bayesian recursive neural network for extended object tracking. Experiment results on simulated and real-world datasets show that the proposed methods outperforms traditional extended object tracking methods and state-of-the-art deep learning approaches.
\end{abstract}


\section{Introduction}
\IEEEPARstart{W}{ith} the improvement of sensor resolution, such as ultra-wideband radar, a target becomes spatially spread scatterers and may occupy multiple resolution cells \cite{ref0}. Therefore, it is possible to jointly estimate both the kinematic states and the extension of the target. This methodology, termed {\em extended object tracking (EOT)}, has seen rapid development over the past decade. It enhances the description of the target by incorporating its size, shape, and orientation, thus providing a more comprehensive characterization of the target. Consequently, EOT has found widespread applications in fields such as autonomous driving \cite{granstrom2023tutorial}, \cite{hirscher2016multiple}, maritime surveillance \cite{vivone2016joint}, and aircraft formation tracking~\cite{liu2018structure}.

Mainstream EOT methods based on Bayesian filtering (BF) framework can be categorized into three distinct types: 1) random matrix models (RMM); 2) random hypersurface models (RHM); and 3) multiplicative error models (MEM). The RMM-based approach was first introduced in \cite{koch2008bayesian}. The target’s extension is represented by a symmetric positive definite (SPD) matrix, which can be equivalently described as an ellipse, referred to as the extension matrix. The extension matrix is modeled as a random matrix obeying an inverse Wishart distribution. Bayesian inference is then employed to predict and update the probability densities of the state and the extension. The RHM-based method for EOT was first introduced in \cite{baum2014extended}, which assumes the measurement source lies on a scaled version of the extension boundaries, to develop Gaussian estimators for EOT with ellipses and free-form star-convex shapes. To maintain the conjugate Gaussian distribution during the evolutionary process, MEM-based methods \cite{yang2019tracking} introduce a multiplicative noise term in the measurement process. These approaches make each scatterer equivalent to the centroid position plus the scaled object size. 

The aforementioned conventional EOT online optimization methods are model-based (MB) methods, namely, they all rely on prior models. Such approaches can theoretically guarantee optimal filtering accuracy when the prior model faithfully captures the environment, whereas the introduction of prior models also introduces additional parameters, which can result in higher filtering errors in practical applications due to model mismatch or parameter mismatch \cite{khodarahmi2023review}. Furthermore, different EOT methods focus solely on redefining the measurement process to achieve better filtering accuracy, while neglecting improvements in the evolution processes of both state and extension (e.g., prediction errors caused by the first-order Markov approximation in the evolutionary process). Additionally, there exists a coupled relationship between the target's state and extension in practical scenarios. To achieve recursive processing, the methods mentioned above typically decouple the target's state and extension roughly during calculations, which significantly limits the accuracy of the filters \cite{wen2024velocityref}.

With the rapid development of deep learning in recent years, an increasing number of deep neural networks (DNNs) have been proposed for real-world tasks. These data-based (DB) methods are typically able to capture the rich features and patterns present in offline data, thereby superseding MB methods in various complex tasks, such as time series analysis~\cite{lim2021time}. Benefiting from the powerful capabilities of recurrent neural network (RNN) architectures for nonlinear features and the ability to extract temporal contextual information, naive RNN and their derivatives, such as gated recurrent units (GRU) \cite{chung2014empirical} and long short-term memory (LSTM) networks \cite{schmidhuber1997long}, have been widely utilized in target tracking \cite{capobianco2021deep}. However, these methods often function as incomprehensible ``black boxes'' to users, with their internal mechanisms being difficult to express analytically, leading to a lack of interpretability. Moreover, since pure neural networks do not incorporate prior models into the prediction process, they exhibit a strong dependency on the quantity of data. Even simple functions are challenging for DNN to learn and understand when data is sparse.

The aforementioned challenges faced in  EOT  can be summarized in the following two points: 1) the errors introduced by insufficient decoupling of state and extension in MB methods, as well as the degraded performance caused by model mismatches in non-Markovian scenarios; 2) the inefficiency in convergence and inference of DB methods due to the lack of priors, which limits their effectiveness in practical tasks. 
Drawing inspiration from the recent success of such metholodogies in realizing hybrid MB/DB algorithms for (non-extended) tracking algorithms, e.g.,~\cite{revach2022kalmannet,ni2022rtsnet,imbiriba2023augmented, buchnik2023latent,gama2023unsupervised, nuri2024learning} (see also survey in~\cite{shlezinger2024ai}), this paper proposes a Bayesian EOT method enhanced by deep learning. Namely,  we utilize DNNs to augment MB Bayesian EOTs, as a form of model-based deep learning~\cite{shlezinger2023model}, aiming to benefit from the interpretability and model-awareness of Bayesian EOTs, alongside the abstractness of DNNs.

Specifically, the contributions of this work are as follows:
\begin{itemize}
    \item A memory mechanism distilled from offline data is proposed to transform the non-Markovian EOT model  into a first-order Markov model, while decoupling the complex relationship between the state and the extension.
    \item Based on the equivalent model, a BF framework composed of memmory is proposed to describe the joint iterative process of memory, state, and extension.
    \item Under Gaussian approximation, a closed-form solution corresponding to the proposed BF framework is derived, achieving efficient recursive computation while providing accurate estimation of both the state and the extension.
    \item A dual-branch memory-aided BF DNN, termed {\em \nameeot}, is proposed. This algorithm is derived from the prior model under the Bayesian framework, providing  interpretability while leverages the DNN's ability to learn and extract nonlinear patterns, granting it offline learning capabilities. Through carefully designed self-supervised training methods, the \nameeot~outperforms various advanced MB EOT methods in benchmark experiments.
\end{itemize}

This paper is organized as follows:  Section \ref{s_2} presents the fundamental theory of the RMM approach and its novel modeling method for real-world scenarios. In Section \ref{s_3} we derive our EOT algorithm by ransforming the proposed model into an equivalent Markov model, followed by proving its closed-form expressions, and  introducing \nameeot~along with its end-to-end training approach. Detailed comparative experimental results are provided in Section \ref{s_4}. Subsequently, Section \ref{s_5} concludes the paper.

Notation: throughout this paper, the superscript $(\cdot)^\mathrm{T}$ denotes the transpose operation; $\mathrm{tr}(\cdot)$ represents calculating the trace of a matrix; $|\cdot|$ represents the calculation of a matrix’s determinant; $\mathbb{E}_{(\cdot|\cdot)}[\cdot]$ denotes the conditional expectation; $(\cdot)_{k|k-1}$ and $(\cdot)_{k|k}$ stand for the prediction and posterior of the state and extension, respectively; $d(\cdot)$ represents calculate the dimension of the corresponding vector. If it is a matrix, the dimension is calculated based on the flattened form of the matrix.

\section{Problem Formulation} \label{s_2}
Building upon the methodology proposed in \cite{lan2014tracking}, in this section we formulate the RMM-based EOT framework, which divides the tracking process into two distinct phases of state prediction and state updates. We first review existing modeling, based on which we introduce our first-order Markov representation.

\subsection{Existing Dynamic Model}

We consider a single-target tracking scenario within a two-dimensional plane. At time instant $k$, the kinematic state of the target's centroid is a $2d \times 1$ random vector $\mathbf{x}_k$ (specifically, for $d=3$, the vector encompasses position, velocity, and acceleration). The target's extension is a $d\times d$ symmetric positive definite (SPD) matrix $\mathbf{X}_k$, which encapsulates its size, shape and orientation.

The state evolution model is constructed as
\begin{gather}
    \mathbf{x}_k = f_k(\mathbf{x}_{k-1}) + \mathbf{w}_k, \ \mathbf{w}_k \sim \mathcal{N}(\mathbf{0}, \mathbf{D}_k \otimes \mathbf{X}_k), \label{state_pred}
\end{gather}
where, $f_k(\cdot)$ represents the nominal (non)linear state-transition function; the process noise $\mathbf{w}_k$  assumed to obey an independent zero-mean Gaussian distribution with covariance of $\mathbf{D}_k \otimes \mathbf{X}_k$, where ``$\otimes$'' signifies the Kronecker product; $\mathbf{D}_k=\sigma_k^2\hat{\mathbf{D}}_k$ is the covariance matrix of the process noise in the one-dimensional model, where $\sigma_k^2$ is the variance of the acceleration noise, $\hat{\mathbf{D}}_k=(1-e^{-2\tau/\theta})\mathrm{diag}([0,0,1])$ is the parametric matrix introduced by \cite{koch2008bayesian}, with $\tau$ representing the temporal interval between sequential frames.

Notably, in Eq. \eqref{state_pred}, the covariance of the process noise $\mathbf{x}_k$ is modeled to reflect dependencies on the target's extension $\mathbf{X}_k$, diverging from the standard Kalman filtering (KF) approach. In standard KF formulations, process noise is generally presumed to encapsulate intrinsic system characteristics, commonly represented in a form such as $\mathbf{Q}_k$, which is a SPD matrix.

Furthermore, compared to the standard KF \cite{lan2014tracking}, the state evolution in Eq. \eqref{state_pred} includes additional descriptions of the extension evolution process as
\begin{gather}
    p(\mathbf{X}_k|\mathbf{X}_{k-1}) = \mathcal{W}(\mathbf{X}_k;\delta_k,\phi_k(\mathbf{X}_{k-1})), \label{extension_pred}
\end{gather}
where $\phi_k(\cdot)$ represents the nominal (non)linear extension-transition functions; $\mathcal{W}(\mathbf{Y};a,\mathbf{C})$ stands for the density of the Wishart distribution for an SPD random matrix $\mathbf{Y}$. Typically, $\phi_k(\cdot) = \mathbf{A}_k(\cdot) \mathbf{A}_k^\mathrm{T}$ \cite{lan2014tracking}, where $\mathbf{A}_k$ can be used to describe changes in the target's extension, such as the orientation (if $\mathbf{A}_k$ is a rotation matrix) or size (if $\mathbf{A}_k = \lambda \mathbf{I}_d$); $\delta_k$ describes the uncertainty of the extension transition \cite{koch2008bayesian}, when $\delta_k$ increases, the precision in delineating the transition process is correspondingly enhanced, and vice versa.

\subsection{Existing Measurement Model}

We use the notation $\mathbf{Z}_k = \{\mathbf{z}_k^i\}_{i=1}^{n_k}$ to represent that the target has yielded $n_k$ measurement vectors $\mathbf{z}_k^i$ during a single measurement. When employing observational sensors (such as radar), $\mathbf{z}_k^i$ stands for the scatterers generated by the target, which predominantly encapsulates the locational attributes of the target. The measurement process is characterized as
\begin{gather}
    \mathbf{z}_k^i = h_k(\mathbf{x}_k) + \mathbf{v}_k^i, \  \mathbf{v}_k^i \sim \mathcal{N}(\mathbf{0}, \mathbf{B}_k \mathbf{X}_{k}\mathbf{B}_k^\mathrm{T}), \label{measurement}
\end{gather}
 where $h_k(\cdot)$ represents the nominal (non)linear measurement function; $\mathbf{v}_k^i$ denotes the measurement noise corresponding to the $i$-th scatter, which is independent of $\mathbf{v}_k^r$ $( i, r = 1, \ldots, n_k, i \neq r )$, and obeys a zero-mean Gaussian distribution with covariance of $\mathbf{B}_k \mathbf{X}_k \mathbf{B}_k^\mathrm{T}$; $\mathbf{B}_k$ is equivalent to the observation distortion matrix as described in \cite{lan2014tracking}, where the distortions in the observation can originate from the geometric relationship between the target and the sensor, making $\mathbf{B}_k$ a function of $\mathbf{x}_k$.

The above measurement model assumes that all scatterers are generated by the centroid of the target and describes the dispersion of the point clouds through the covariance of the measurement noise. Additionally, the measurement error inherent to the sensor itself can be equivalently represented as the noise term in Eq. \eqref{measurement} \cite{feldmann2008tracking}. When $n_k=1$, the measurement process with an extended form reduces to the form used in point object tracking tasks, as employed by the standard KF. In this case, the covariance of the measurement noise is typically assumed to take a form $\mathbf{R}_k$.

In \cite{koch2008bayesian}, it is assumed that the number of scatterers $n_k$ is independent of both $\mathbf{x}_k$ and $\mathbf{X}_k$. Consequently, the likelihood of obtaining the measurement $\mathbf{Z}_k$, given the state, the extension, and the number of scatterers, takes the form of a product of $n_k$ Gaussian distributions, namely,
\vspace{-0.5em}
\begin{gather}
    p(\mathbf{Z}_k|n_k,\mathbf{x}_k,\mathbf{X}_k) = \prod_{i=1}^{n_k}\mathcal{N}(\mathbf{z}_k^i;h_k(\mathbf{x}_k),\mathbf{B}_k \mathbf{X}_{k}\mathbf{B}_k^\mathrm{T}) \nonumber \\
    \propto \! \mathcal{N}(\widetilde{\mathbf{z}}_k;h_k(\mathbf{x}_k),\frac{\mathbf{B}_k \mathbf{X}_{k}\mathbf{B}_k^\mathrm{T}}{n_k})  \mathcal{W}(\widetilde{\mathbf{Z}}_k;n_k-1,\mathbf{B}_k\mathbf{X}_k\mathbf{B}_k^\mathrm{T}), 
\end{gather} 
\vspace{-0.7em}
where 
\begin{align}
    \widetilde{\mathbf{z}}_k = \frac{1}{n_k}\sum_{i=1}^{n_k}\mathbf{z}_k^i, \ \ \ \widetilde{\mathbf{Z}}_k=\sum_{i=1}^{n_k}(\mathbf{z}_k^i-\widetilde{\mathbf{z}}_k)(\mathbf{z}_k^i-\widetilde{\mathbf{z}}_k)^\mathrm{T}.  
\end{align}

\subsection{Flaw of RMM-Based EOT}

Both Eqs. \eqref{state_pred} and \eqref{extension_pred} employ the first-order Markov assumption to characterize the relationship between two consecutive frames in the dynamic evolution process. This is equivalent to the prevalent assumption within the State Space Model (SSM) that the contribution of distant historical frames to the current state space is negligible compared to that of the immediately preceding frame~\cite{durbin2012time}. The introduction of this hypothesis simplifies the analysis of the dynamic evolution process and allows it to be computed recursively and efficiently. These advantages have led to the widespread application of the first-order Markov assumption in RMM-based EOT methods.

However, the first-order Markov assumption introduces error: discard all historical state information except the preceding frame significantly increases the risk of losing critical information. This can result in substantial biases in the estimation of both the target's state and its extension.

\begin{figure}
    \centering
    \includegraphics[width=0.8\linewidth]{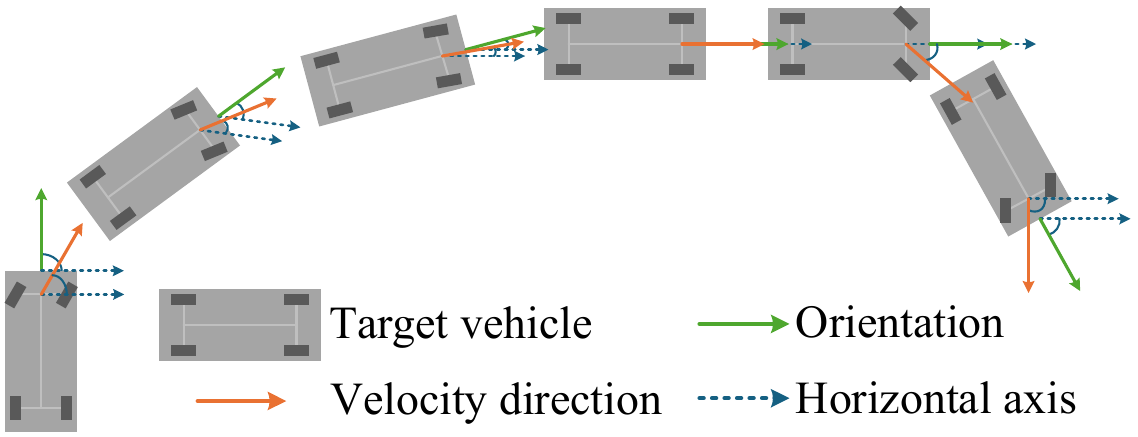}
    \caption{Illustration of the non-Markovian dynamics and the complex coupling between state and extent in real-world scenarios.\protect}
    \label{non-Markovian proprety in real world}
\end{figure}

According to the hypotheses outlined in {\cite{koch2008bayesian}}, the target's state $\mathbf{x}_k$ exhibits an explicit dependence on the extension $\mathbf{X}_k$; that is, the evolution of the centroid state is influenced by the morphology of the extension as decribed in Eq. \eqref{state_pred}, while the evolution of the extension is independent of the target's state. This assumption appears to be counterintuitive, as in practical EOT tasks, the state and extension often exhibit intricate interdependencies, (see Fig. \ref{non-Markovian proprety in real world}). The deviation causes the state to exert an influence on the evolution of the extension (for example, the change rate of the orientation of a turning vehicle is directly proportional to its angular velocity of its wheels). Consequently, the previously assumed independence between state and extension is no longer applicable.

The measurement model for EOT, as defined in Eq. \eqref{measurement}, is only an approximation to the measurement process. As such, it can present substantial discrepancies with the actual measurement processes in real-world tasks. For instance, improperly calibrated sensors can introduce unquantifiable biases in the measurements. In addition, in the measurement process of the real world, the geometric relationship between the target's position, orientation angle, and the sensor often affects the measurement equation. Consequently, the measurement matrix $h_k(\cdot)$ in Eq. \eqref{measurement}, must be expressed as a function of $\mathbf{x}_k$ and $\mathbf{X}_k$. This factor is frequently neglected in conventional modeling approaches. 

\subsection{Proposed Model}

Inspired by  \cite{gonzalez2021learning}, we  develop a respective state and extension evolution model with non-Markovian properties to remedy the estimation errors introduced by the first-order Markov approximation. This model couples the state and extension to better capture the complexities of real-world evolution processes. The specific equations are formulated as 
\begin{flalign}
    & \emph{State evolution:}& \nonumber \\
    & \quad \quad \mathbf{x}_k = f_k^{\mathrm{general}}(\mathbf{x}_{k-1}, ..., \mathbf{x}_1, \mathbf{X}_{k-1}, ..., \mathbf{X}_{1}) + \mathbf{w}_k; \label{general_state_pred} \\
    & \emph{Extension evolution:}& \nonumber \\
    & \quad p(\mathbf{X}_k|\mathbf{x}_{k-1}, ..., \mathbf{x}_1, \mathbf{X}_{k-1}, ..., \mathbf{X}_{1})= \nonumber \\
    & \quad \quad \quad \mathcal{W}(\mathbf{X}_k;\delta_k,\phi_k^{\mathrm{general}}(\mathbf{x}_{k-1}, ..., \mathbf{x}_1, \mathbf{X}_{k-1}, ..., \mathbf{X}_{1}));  \label{general_extension_pred}
\end{flalign} 
\noindent where $f_k^{\mathrm{general}}(\cdot)$ and $\phi_k^{\mathrm{general}}(\cdot)$ denote the generalized state and extension evolution function, respectively. These functions encapsulate the entirety of the historical state and extension data into their respective evolutionary frameworks, thereby providing a thorough adaptation to real-world evolution models. However, this comprehensive modeling results in increased dimensionality, which complicates the modelling of their PDFs. It is noteworthy that when Eqs. \eqref{general_state_pred} and \eqref{general_extension_pred} are reduced to include only $\mathbf{x}_k$ and $\mathbf{X}_k$, respectively, they simplify to the single frame forms delineated in Eqs. \eqref{state_pred} and \eqref{extension_pred}. 

 Similarly, the general measurement model for extended targets can be defined as:
\begin{equation}
    \mathbf{z}_k^i = h_k^{\mathrm{general}}(\mathbf{x}_k, \mathbf{X}_k) + \mathbf{v}_k^i, \label{general_measurement}
\end{equation}
where $h_k^{\mathrm{general}}(\cdot)$ is the generalized measurement function.

The above model can faithfully describe EOT settings, overcoming some of the limitations induced by conventional first-order Markov approximations. However, due to the inherent high-dimensional evolution characteristics of the non-Markovian framework mentioned above, its application in real-time tracking tasks becomes complex, making it challenging to perform recursion through heuristic methods. Thus, we need to develop a non-Markovian framework with efficient recursive properties.

In summary, to effectively address the aforementioned challenges, our goal is to establish a hybrid approach that combines model priors with offline data for the joint estimation problem of non-Markovian states and extensions under uncertainty. Drawing inspiration from \cite{yan2024explainable}, an efficient explainable Bayesian framework is proposed, which integrates state, extension, and memory, achieving precise target estimation while maintaining efficient recursive performance.

\section{Deep Memory Aided Bayesian Filter for EOT} \label{s_3}
This section elucidates the design process of the framework in three sequential steps. First, we transform the non-Markovian model into an relaxed first-order Markovian model by compensating for the errors in both the evolution and measurement processes. Then, based on the relaxed  Markovian model, an interpretable Bayesian filtering framework is devised, integrating state, extension, and memory. 
We proceed to provide a feasible implementation of the filtering framework based on Gaussian approximations and moment matching methods. For specific variables in the filtering process that lack closed-form analytical expressions, deep learning methods are leveraged to construct them, with the aim of deriving iterative analytical expressions. To that end, our proposed {\em \nameeot}, a MB dual branch memory DNN, is designed.

\subsection{Relaxed Model}
\subsubsection{Transformed First-Order Model}
The non-Markovian evolution processes in Eqs. \eqref{general_state_pred} and \eqref{general_extension_pred} are characterized by high dimensional complexity, thereby rendering them infeasible for recursive computation. By introducing evolution functions $f_k(\cdot)$ and $\phi_k(\cdot)$, which exhibit first-order Markovian properties into the aforementioned models, evolution equations described by process mismatch terms are derived in Eqs. \eqref{state_pred_mismatch} and \eqref{extension_pred_mismatch} (shown at the top of the next page), 
\begin{figure*}[t!]
    \vspace*{-1em}
    \begin{gather}
        \mathbf{x}_k = \underbrace{f_k^{\mathrm{general}}(\mathbf{x}_{k-1},...,\mathbf{x}_{1}, \mathbf{X}_{k-1}, \mathbf{X}_{1}) - f_k(\mathbf{x}_{k-1})}_{\bm{\Delta}_k^f} + f_k(\mathbf{x}_{k-1}) + \mathbf{w}_k .\label{state_pred_mismatch} \\
        p(\mathbf{X}_k|\mathbf{x}_{k-1}, ..., \mathbf{x}_1, \mathbf{X}_{k-1}, ..., \mathbf{X}_{1}) = \mathcal{W}(\mathbf{X}_k;\delta_k,\underbrace{\phi_k^{\mathrm{general}}(\mathbf{x}_{k-1},...,\mathbf{x}_{1}, \mathbf{X}_{k-1}, \mathbf{X}_{1}) - \phi_k(\mathbf{X}_{k-1})}_{\bm{\Delta}_k^\phi}+ \phi_k(\mathbf{X}_{k-1})) . \label{extension_pred_mismatch}
    \end{gather}
    \hrulefill
    \vspace*{-3mm}
\end{figure*}
where $\bm{\Delta}_k^f$ and $\bm{\Delta}_k^\phi$ stand for the model mismatch errors in the state and extension evolution processes, respectively.

Similarly, by introducing the nominal measurement function $h_k(\cdot)$, the measurement equation incorporating the measurement mismatch error can be defined as:
\begin{gather}
\mathbf{z}_k^i = \underbrace{h_k^{\mathrm{general}}(\mathbf{x}_k, \mathbf{X}_k) - h_k(\mathbf{x}_k)}_{\bm{\Delta}_k^h} + h_k(\mathbf{x}_k) + \mathbf{v}_k^i, \label{measurement_mismatch}
\end{gather}
where $\bm{\Delta}_k^h$ stands for the model mismatch errors in the measurement.
In Eq. \eqref{measurement_mismatch}, the variability among scatterers within a frame depends solely on the measurement noise $\mathbf{v}_k^i$ and is independent of $\bm{\Delta}_k^h$. Therefore, the model mismatch error in the measurement remains constant over the $n_k$ observations.

Owing to the incorporation of first-order Markovian approximation based transition functions in Eqs. \eqref{state_pred_mismatch} and \eqref{extension_pred_mismatch}, the resultant model mismatch errors are composed of the cumulative historical states and extension, thereby complicating their representation through analytical expressions. To simplify the equations motioned above, the evolution mismatch errors $\bm{\Delta}_k^f$ and $\bm{\Delta}_k^\phi$ can be articulated using a hierarchical nested function:
\begin{gather}
    \mathbf{c}_k = g_k^\Delta(\mathbf{x}_{k-1}, ..., \mathbf{x}_1, \mathbf{X}_{k-1}, ..., \mathbf{X}_1), \label{c_k_update} \\
    \bm{\Delta}_k^f = f_k^\Delta(\mathbf{c}_k), \ \bm{\Delta}_k^{\phi} = \phi_k^\Delta(\mathbf{c}_k). \label{generate_delta_k_phi}
\end{gather}

Note that $\mathbf{c}_k$ denotes a memory term, encapsulating the state and extension patterns from time steps $1$ to $k-1$, thus exhibiting non-Markovian properties; the nonlinear function $g_k^\Delta(\cdot)$ is utilized to map high-dimensional historical features into the latent space, whereas $f_k^\Delta(\cdot)$ and $\phi_k^\Delta(\cdot)$ signify the nonlinear state and extension output functions, respectively, which map the latent memory information to the expressible state and extension evolution model mismatch terms.

The feature embedding function $g_k^\Delta(\cdot)$ in Eq. \eqref{c_k_update} lacks a concrete physical interpretation. Nevertheless, and consistent with the approach in \cite{yan2024explainable}, we postulate that $g_k^\Delta(\cdot)$ adheres to a fixed pattern, signifying its temporal invariance, i.e., $g_k^\Delta(\cdot) \Leftrightarrow g^\Delta(\cdot)$. Therefore, the memory update process can be described in a heuristic manner as follows:
\begin{align}
    \mathbf{c}_k & = \underbrace{g^c(g^c(g^c}_{k \ \mathrm{times}}(\cdots), \mathbf{x}_{k-2}, \mathbf{X}_{k-2}), \mathbf{x}_{k-1}, \mathbf{X}_{k-1}) \nonumber \\
    & = g^c (\mathbf{c}_{k-1}, \mathbf{x}_{k-1}, \mathbf{X}_{k-1}). \label{c_k_update_iter}
\end{align}

Similarly, the error term $\bm{\Delta}_k^h$ arising from measurement model mismatches  can be mapped through the nonlinear measurement output function $h^\Delta(\cdot)$, namely:
\begin{equation}
    \bm{\Delta}_k^h = h^{\Delta}(\mathbf{x}_k, \mathbf{X}_k). \label{generate_delta_k_h}
\end{equation}

By transforming the aforementioned problem, we have demonstrated that the joint state-extension evolution model, characterized by non-Markovian properties, can be equivalently reduced to an iterative first-order Markov model by incorporating a memory mechanism. Nevertheless, the complex coupling among memory, state, and extension in Eq. 
\eqref{c_k_update_iter} complicates the memory update process, making it difficult to articulate through concise analytical expressions.

\subsubsection{Model Identification}
Inspired by data-driven approaches for system identification~\cite{chiuso2019system, ljung2020deep}, we propose approximating the memory term by learning from offline data. Given that target motion generally exhibits analogous state evolution, extension evolution, and measurement, deep learning methods suitable for time series prediction can be employed to effectively distill common patterns from offline data. This approach can supplant the estimation process of $\mathbf{c}_k$, thereby improving estimation accuracy during the evolution process. 

Let $\mathcal{D}$ denote the offline historical data, encompassing time-varying target states, true extents, and corresponding measurement sequences, i.e.,
\begin{equation}
    \mathcal{D}=\{\bar{\mathbf{x}}_{1:K}^j, \overline{\mathbf{X}}_{1:K}^j, \mathbf{Z}_{1:K}^j \}_{j=1}^J, \label{offline_dataset}
\end{equation}
\noindent where $\bar{\mathbf{x}}_{1:K}$, $\overline{\mathbf{X}}_{1:K}$, and $\mathbf{Z}_{1:K}$ denote the ground truth sequence of target's states, extension, and corresponding measurements, respectively; $K$ represents the length of each sequence, while $J$ indicates the number of sequence in the offline dataset. Such an offline dataset can be generated through straightforward simulations, for instance, by collecting the motion trajectories of a vehicle and the temporal variations in its extension.

With the introduction of learnable offline data, the problem of estimating model mismatch errors is transformed into the task of learning the mapping functions $f_k^\Delta(\cdot)$, $\phi_k^\Delta(\cdot)$, and $h_k^\Delta(\cdot)$ from the offline dataset $\mathcal{D}$. However, in contrast to traditional Bayesian methods that naively estimate only the state and extension, the incorporation of memory and model mismatch errors complicates the evolution process, thus necessitating recursive expressions with analytical forms.

\subsection{Bayesian Filtering Framework Design}
Before deriving analytical expressions for the filter, we establish a theoretically feasible Bayesian framework to describe the transition of probability densities. The Bayesian framework consists of the following two steps: prediction and update of the probability density of the estimated variables. In Theorem \ref{T_1}, we integrated the memory term $\mathbf{c}_k$ and offline data $\mathcal{D}$ into the Bayesian framework of the conventional RMM, resulting in an extended Bayesian framework that incorporates state-extension-memory mismatch error. However, these formulas are challenging to apply directly in practical tasks. Therefore, in Theorem \ref{T_2} we employ a Gaussian assumption and moment-matching approach to derive an analytical expression for the probability density integrals process in Theorem \ref{T_1}, this theorem is derivation of the joint probability density presented in \cite{yan2024explainable}. These expressions are applicable to the filter under the model given by Eqs. \eqref{general_state_pred} - \eqref{general_measurement}.

\begin{theorem}\label{T_1}{\textsl{(Joint state-extension-memory Bayesian filtering framework)}}
    For the state, extension and memory evolution in Eqs. (\ref{state_pred_mismatch}) - (\ref{generate_delta_k_h}), given the previous measurements $\mathbf{z}_{1:k-1}$, the offline dataset $\mathcal{D}$, and the previous joint state-extension-memory-mismatch posterior density $p(\mathbf{x}_{k-1}, \mathbf{X}_{k-1}, \mathbf{c}_{k-1}|\mathbf{Z}_{1:k-1}, \mathcal{D})$, the joint state-extension-memory density prediction is
    \begin{align}
        & \!\!\!\!\!\!\! p(\mathbf{x}_k,\mathbf{X}_k,\mathbf{c}_k|\mathbf{Z}_{1:k-1},\mathcal{D}) = \nonumber \\ 
        & \!\!\!\!\!\!\!\!\!\! \int \!\!\!\!\! \int \!\!\!\!\! \int \!\!\! P_k^1 p(\mathbf{x}_{k-1}, \mathbf{X}_{k-1}, \mathbf{c}_{k-1}|\mathbf{Z}_{1:k-1}, \mathcal{D}) \mathrm{d}\mathbf{x}_{k-1}\mathrm{d}\mathbf{X}_{k-1}\mathrm{d}\mathbf{c}_{k-1}. \label{aabbcc}
    \end{align}
    with 
    \begin{align}
        P_k^1 &= \iint p(\mathbf{x}_k|\bm{\Delta}_k^f,\mathbf{x}_{k-1},\mathbf{X}_k) p(\bm{\Delta}_k^f|\mathbf{c}_k, \mathcal{D}) \nonumber \\
        & \quad\quad \times p(\mathbf{X}_k|\bm{\Delta}_k^\phi,\mathbf{X}_{k-1}) p(\bm{\Delta}_k^\phi|\mathbf{c}_k,\mathcal{D}) \nonumber \\
        & \quad\quad \times p(\mathbf{c}_k|\mathbf{x}_{k-1},\mathbf{X}_{k-1},\mathbf{c}_{k-1},\mathcal{D})\mathrm{d}\bm{\Delta}_k^f\mathrm{d}\bm{\Delta}_k^\phi. \label{aacc}
    \end{align}
    The joint state-extension-memory density update is
    \begin{align}
        & p(\mathbf{x}_k, \mathbf{X}_k, \mathbf{c}_k|\mathbf{Z}_{1:k}, \mathcal{D}) =  
        p(\mathbf{x}_k, \mathbf{X}_k, \mathbf{c}_k|\mathbf{Z}_{1:k-1}, \mathcal{D}) \nonumber\\ 
        & \quad \quad \quad \times \frac{\int p(\mathbf{Z}_k|\bm{\Delta}_k^h,\mathbf{x}_k,\mathbf{X}_k)p(\bm{\Delta}_k^h|\mathbf{x}_k,\mathbf{X}_k,\mathcal{D})\mathrm{d}\bm{\Delta}_k^h}{p(\mathbf{Z}_k|\mathbf{Z}_{1:k-1}, \mathcal{D})} .
    \end{align}
\end{theorem}

The iteration process of the joint state-extension-memory BF contains three procedures: 1) the memory update process determines the current memory based on the previous state, previous extension, previous memory, and offline data; 2) the joint prediction process determines the joint state-extension-memory prediction PDF; 3) the joint update process determines the joint update PDF of the state, extension and memory. 

Although the aforementioned BF framework appears to possess a complete iterative process, which could potentially lead to the derivation of a solution in a closed-form expression. However, the following probability densities, including the memory update density $p(\mathbf{c}_k|\mathbf{x}_{k-1}\mathbf{X}_{k-1},\mathbf{c}_{k-1},\mathcal{D})$, the state evolution mismatch density $p(\bm{\Delta}_k^f|\mathbf{c}_k,\mathcal{D})$, the extension evolution mismatch density $p(\bm{\Delta}_k^\phi|\mathbf{c}_k,\mathcal{D})$, and the observation mismatch density $p(\bm{\Delta}_k^h|\mathbf{x}_k,\mathbf{X}_k,\mathcal{D})$, are generally difficult to model, rendering the derivation of iterative analytical expressions challenging.

Fortunately, the information inherent in rich offline datasets encapsulates the characteristics of these PDFs. Consequently, designing appropriate DNN modules may allow for the learning of the mapping of these unknown distributions from offline data. Once the forms of these  PDFs have been ascertained, the analytical implementation of the filtering can be derived independently.

Before designing the specific DNN modules, it is imperative to derive the closed-form solution of the Bayesian filtering framework as stated in Theorem \ref{T_1}. The state filtering process of the proposed framework has various implementation methods, such as Monte Carlo sampling \cite{el2021policy} and Gaussian approximation \cite{wang2012gaussian}. Since Monte Carlo sampling typically optimizes the estimated values of variables in an online manner, it is often challenging to meet the low-latency requirements of real-time systems. In contrast, expressions derived from Gaussian approximation generally exhibit favorable analytical properties, making them computationally efficient and thus consuming fewer system resources. For this reason, we employ Gaussian approximation with moment matching to derive the state filtering expressions. The following outlines the assumptions necessary to apply the Gaussian approximation.

\noindent \textbf{Assumption 1.} The state transition probability density at time $k$ obeys a Gaussian distribution with a mean of $f_k(\mathbf{x}_{k-1})+\bm{\Delta}_k^f$ and a covariance of $\mathbf{Q}_k$; and the extension transition probability density obeys a Wishart density with scaler parameter and parameter matrix of $\delta_k$ and $\mathbf{A}_k\mathbf{X}_{k-1}\mathbf{A}_k^\mathrm{T} + \bm{\Delta}_k^\phi$. Those conditional densities  are given as
\begin{gather}
    p(\mathbf{x}_k|\bm{\Delta}_k^f,\mathbf{x}_{k-1},\mathbf{X}_k) = \mathcal{N}(\mathbf{x}_k;f_k(\mathbf{x}_{k-1})+\bm{\Delta}_k^f, \mathbf{Q}_k), \label{asdvasdfv} \\ 
    p(\mathbf{X}_k|\bm{\Delta}_k^\phi,\mathbf{X}_{k-1}) = \mathcal{W}(\mathbf{X}_k;\delta_k,\mathbf{A}_k\mathbf{X}_k\mathbf{A}_k^\mathrm{T}+\bm{\Delta}_k^\phi). \label{asdasd}
\end{gather}
\noindent \textbf{Assumption 2.} The memory update PDF obeys a Gaussian distribution with a mean of $\hat{\mathbf{c}}_k$ and a covariance of $\mathbf{P}_k^c$, namely,
\begin{align}
    p(\mathbf{c}_k|\mathbf{x}_{k-1},\mathbf{X}_{k-1},\mathbf{c}_{k-1},\mathcal{D}) = \mathcal{N}(\mathbf{c}_k;\hat{\mathbf{c}}_k,\mathbf{P}_k^c) .\label{memoryasdasd}
\end{align}
\noindent \textbf{Assumption 3.} 
$\bm{\Delta}_k^f$ obeys a Gaussian distribution with expectation and covariance of $\hat{\bm{\Delta}}^f_k$ and $\mathbf{P}_k^f$, respectively; and $\bm{\Delta}_k^h$ obeys a Gaussian distribution with expectation and covariance of $\hat{\bm{\Delta}}h_k$ and $\mathbf{P}_k^h$, respectively. The densities motioned above are given as
\begin{gather}
    p(\bm{\Delta}_k^f | \mathbf{c}_k, \mathcal{D}) = \mathcal{N}(\bm{\Delta}_k^f;\hat{\bm{\Delta}}^f_k, \mathbf{P}_k^f) ,\label{delta_f} \\
    p(\bm{\Delta}_k^h | \mathbf{x}_k, \mathbf{X}_k, \mathcal{D}) = \mathcal{N}(\bm{\Delta}_k^h;\hat{\bm{\Delta}}^h_k, \mathbf{P}_k^h). \label{delta_h}
\end{gather}

Based on Assumptions 1-3, we propose to DNNs to lean the needed features for their computation from data. Specifically,
the first- and second-order moments corresponding to the aforementioned model mismatch terms are regressed through a linear layer. Specifically, $\hat{\bm{\Delta}}^f_k$ and $\mathbf{P}_k^f$ are obtained from the regression layers parameterized by $\bm{\psi}^f$ and $\bm{\psi}^{\mathbf{P}^f}$, denoted as $\Upsilon_{\bm{\psi}^f}: \mathbb{R}^{d(\hat{\mathbf{c}}_k) + d(\mathbf{P}_k^c)} \mapsto \mathbb{R}^{d(\hat{\bm{\Delta}}^f_k)}$ and $\Upsilon_{\bm{\psi}^{\mathbf{P}^f}}: \mathbb{R}^{d(\hat{\mathbf{c}}_k) + d(\mathbf{P}_k^c)} \mapsto \mathbb{R}^{d(\mathbf{P}^f)}$, respectively. These two linear layers map the joint vector of $\hat{\mathbf{c}}_k$ and $\mathbf{P}_k^c$ to produce the two moments. Similarly, $\hat{\bm{\Delta}}^h_k$ and $\hat{\bm{\Delta}}^h_k$ are derived from regression layers parameterized by $\bm{\psi}^h$ and $\bm{\psi}^{\mathbf{P}^h}$, denoted as $\Upsilon_{\bm{\psi}^h}: \mathbb{R}^{d(\mathbf{x}_{k}) + d(\mathbf{X}_{k})} \mapsto \mathbb{R}^{d(\hat{\bm{\Delta}}^h_k)}$ and $\Upsilon_{\bm{\psi}^{\mathbf{P}^h}}: \mathbb{R}^{d(\mathbf{x}_{k}) + d(\mathbf{X}_{k})} \mapsto \mathbb{R}^{d(\mathbf{P}^h)}$, respectively. All the aforementioned mapping functions are learned from offline data .

\noindent \textbf{Assumption 4.} $\bm{\Delta}_k^\phi$ as an extension evolution mismatch term. It satisfies the conditions that its first moment equals $\mathbf{P}_k^\phi$, namely,
\begin{gather}
    \mathbb{E}_{\bm{\Delta}_k^\phi | \mathbf{c}_k, \mathcal{D}}[\bm{\Delta}_k^\phi] = \mathbf{P}_k^\phi. \label{ex_pred_mis}
\end{gather}

The term $\mathbf{P}_k^\phi$ is generated by mapping the joint vector $\hat{\mathbf{c}}_k$ and $\mathbf{P}_k^c$ through a linear layer parameterized by $\bm{\psi}^{\mathbf{P}^\phi}$, denoted as $\Upsilon_{\bm{\psi}^{\mathbf{P}^\phi}}: \mathbb{R}^{d(\hat{\mathbf{c}}_k) + d(\mathbf{P}_k^c)} \mapsto \mathbb{R}^{d(\hat{\bm{\Delta}}^f_k)}$.\\
\noindent \textbf{Assumption 5.} The state posterior probability density obeys a Gaussian distribution with a mean of $\mathbf{x}_{k|k}$ and a variance of $\mathbf{P}_{k|k} \otimes \mathbf{X}_k$; and the extension posterior probability density follows an inverse Wishart distribution with a scaler parameter of $v_{k|k}$ and a parameter matrix of $\mathbf{X}_{k|k}$, namely
\begin{gather}
    p(\mathbf{x}_k|\mathbf{Z}_{1:k},\mathcal{D})=\mathcal{N}(\mathbf{x}_k;\mathbf{x}_{k|k},\mathbf{P}_{k|k}), \\
    p(\mathbf{X}_k|\mathbf{Z}_{1:k},\mathcal{D}) = \mathcal{IW}(\mathbf{X}_k;v_{k|k},\mathbf{X}_{k|k}). \label{dasdas}
\end{gather}
\noindent \textbf{Assumption 6.} The likelihood to get the measurement $\mathbf{Z}_k$ given both state and extension as well as the number of measurements, follows the form of a product of  $n_k$ Gaussian distributions.
\begin{align}
    p(\mathbf{Z}_k|n_k, \bm{\Delta}_k^h,&\mathbf{x}_k,\mathbf{X}_k) = \nonumber \\
    & \prod_{i=1}^{n_k}\mathcal{N}(\mathbf{z}_k^i;h_k(\mathbf{x}_k)+\bm{\Delta}_k^h,\mathbf{B}_k \mathbf{X}_{k}\mathbf{B}_k^\mathrm{T}). \label{likelihood}
\end{align}

We have substituted all probability densities from the assumptions above into Theorem \ref{T_1}, utilizing moment matching to derive an analytical expression for the filtering, as presented in Theorem \ref{T_2}. Notably, we have circumvented the complex integration process in Theorem \ref{T_1} to obtain these expressions, which possess favorable recursive properties, making them applicable to practical filtering processes.
\begin{theorem}\label{T_2}{\textsl{(Implementation the estimation process for state and extension)}}
    Under assumptions of 1-6, the expressions used for state and covariance prediction are
    \begin{gather}
        \mathbf{x}_{k|k-1} = f_k(\mathbf{x}_{k-1|k-1}) + \underset{\mathrm{learned}}{\underline{\hat{\bm{\Delta}}_k^f}} ,\label{state_pred_final} \\
        \mathbf{P}_{k|k-1} = \mathbf{F}_k\mathbf{P}_{k-1|k-1}\mathbf{F}_k^\mathrm{T} + \mathbf{Q}_k + \underset{\mathrm{learned}}{\underline{\mathbf{P}_k^f}} ,\label{state_cov_pred_final}
    \end{gather}
    \noindent and the expressions for state and covariance update are
    \begin{gather}
        \mathbf{x}_{k|k} = \mathbf{x}_{k|k-1} + \mathbf{P}^{xz}_{k|k-1}(\mathbf{P}^{zz}_{k|k-1})^{-1}(\widetilde{\mathbf{z}}_k - \mathbf{z}_{k|k-1}), \label{state_update_final} \\
    \mathbf{P}_{k|k} = \mathbf{P}_{k|k-1} - \mathbf{P}_{k|k-1}^{xz}(\mathbf{P}^{zz}_{k|k-1})^{-1}(\mathbf{P}_{k|k-1}^{xz})^\mathrm{T}, \label{state_cov_update_final}
    \end{gather}
    \vspace{-1em}
    \noindent where
    \begin{gather}
        \mathbf{z}_{k|k-1} = h_k(\mathbf{x}_{k|k-1}) + \underset{\mathrm{learned}}{\underline{\hat{\bm{\Delta}}_k^h}} ,\label{state_pred_meas} \\
        \mathbf{P}_{k|k-1}^{xz} = \mathbf{P}_{k|k-1} \mathbf{H}_k^\mathrm{T}, \label{P_xz} \\
        \mathbf{P}_{k|k-1}^{zz} = \mathbf{H}_k\mathbf{P}_{k|k-1}\mathbf{H}_k^\mathrm{T} + \frac{\mathbf{B}_k \mathbf{X}_{k}\mathbf{B}_k^\mathrm{T}}{n_k} + \underset{\mathrm{learned}}{\underline{\mathbf{P}_k^h}} .\label{P_zz}
    \end{gather}

    The prediction process for the extension and its uncertainties is expressed as
    \begin{gather}
        \!\!\!\!\!\! v_{k|k-1} = \frac{2\delta_k(\lambda_{k-1} + 1)(\lambda_{k-1} - 1)(\lambda_{k-1} -2)}{\lambda_{k-1}^2(\lambda_{k-1}+\delta_k)} + 2d + 4, \label{v_pred_final} \\
        \!\!\!\! \mathbf{X}_{k|k-1} = \frac{v_{k|k-1}-2d-2}{\lambda_{k-1}}(\delta_k\mathbf{A}_k\mathbf{X}_{k-1|k-1}\mathbf{A}_k^\mathrm{T} + \underset{\mathrm{learned}}{\underline{\mathbf{P}_k^\phi}}) ,\label{extension_pred_final}
    \end{gather}
    and the update process for the extension and its uncertainties is expressed as
    \begin{flalign}
        & \quad \quad \quad \quad \quad v_{k|k} = v_{k|k-1} + n_k ,\label{v_update_final} \\
        \mathbf{X}_{k|k} &= \mathbf{X}_{k|k-1} + \mathbf{S}_k^{-1}(\widetilde{\mathbf{z}}_k-\mathbf{z}_{k|k-1})(\widetilde{\mathbf{z}}_k-\mathbf{z}_{k|k-1})^\mathrm{T} \nonumber \\
        & + \mathbf{B}_k^{-1}\widetilde{\mathbf{Z}}_k\mathbf{B}_k^{-\mathrm{T}}\mathbf{X}_{k|k-1}^{-1} ,\label{extension_update_final}  
    \end{flalign}
    where
    \begin{gather}
        \lambda_{k-1} = v_{k-1|k-1} - 2d - 2 ,\label{lambda_pred_final} \\
        \mathbf{S}_k^{-1} = (\mathbf{H}_k\mathbf{P}_{k|k-1}\mathbf{H}_k^\mathrm{T})\mathbf{X}_{k|k-1}^{-1} + \frac{|\mathbf{B}_k|^{d/2}}{n_k} + \underset{\mathrm{learned}}{\underline{\mathbf{P}_k^h}} .\label{S_update_final}
    \end{gather}
\end{theorem}
\begin{proof}
    See Appendix.
    \renewcommand{\qedsymbol}{}
\end{proof} 

Adopting the Gaussian approximation above, we derive a formally iterative expression for the filter. However, the compensation terms derived from memory in the expression (variables underlined and annotated with ``learned") are challenging to construct. To overcome this limitation, we next design a DNN based on the proposed BF architecture to learn the mapping of errors induced by model mismatches from offline data, thereby endowing the analytical expression of the filter with practical iterability.

\subsection{\nameeot~Design}

\begin{figure*}[ht]
  \centering
  \includegraphics[width=0.9\linewidth]{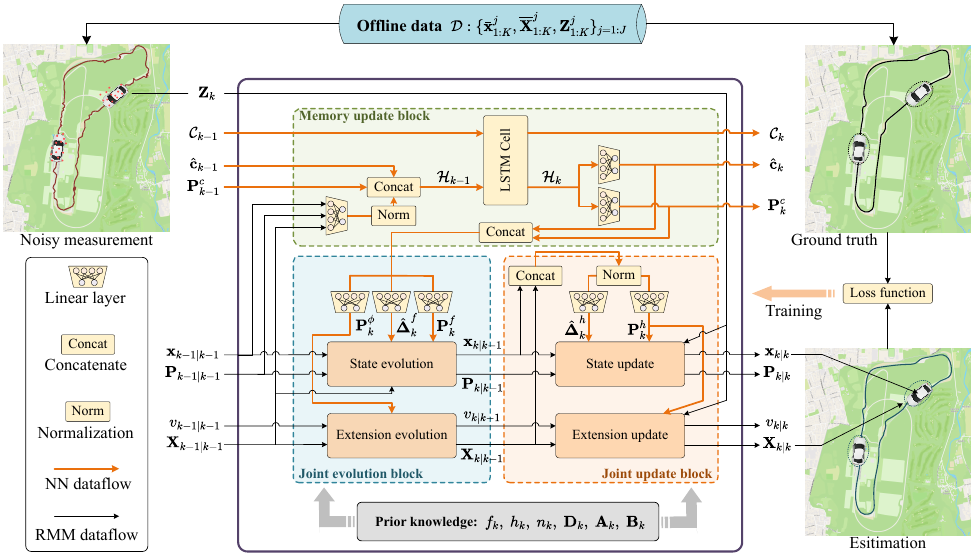}
  \caption{The overall structure of \nameeot.  ``State evolution", ``State update", ``Extension evolution", and ``Extension update" represent the formulas in Theorem \ref{T_2} that exclude the error mismatch terms, derived from the basic RMM. The flow of variables is depicted by black lines. The network parameters along the orange lines (used for generate compensation terms in Theorem \ref{T_2}) are learned via training. }\protect
  \label{eee}
\end{figure*}

The joint state-extension-memory BF shares some similar iterative forms with RNNs, as both are implemented via a step-by-step recursive method over time. Drawing inspiration from the recurrent recursion structure of RNNs, \nameeot~utilizes LSTM cells as its backbone to capture and preserve memory information within sequences in the memory update gate. 
The overall structure of \nameeot, shown in Fig. \ref{eee}, contains DNN modules for estimating the first- and second-order moments of memory and model mismatch errors. 

More precisely, the transmission of memory flow is structured into two branches: $\mathcal{C}_k$, the cell memory flow at time $k$, which is responsible for storing and filtering long-term contextual information, and $\mathcal{H}_k$, the hidden memory, which at each update step first integrates information from all posteriors at time $k\!-\!1$, and then interacts with $\mathcal{C}_k$ to distill more efficient memory features.
In the joint prediction and the joint update gate, we independently utilize a two-layer stacked fully connected layer to select out the useful memory information from the LSTM, which is aimed at generating compensations for the model mismatch errors present in both gates.
Prior information of kinematic and extension is integrated through the filtering equations, and the neural network training relies on offline data. This structure is strictly grounded in BF theory, allowing the network modules in each block to remain simple, without requiring overly complex designs.

Compared to the naive application of RNNs and LSTMs in target tracking, \nameeot~is derived from BF theory, where DNNs are utilized to compensate for model mismatch errors within the filtering process. From the perspective of DNNs, this approach incorporates prior knowledge of the actual physical process, thereby reducing the network's reliance on large datasets. Additionally, the incorporation of second-order moments, representing uncertainty within the BF framework, endows the predictions of \nameeot~with greater interpretability.

\subsection{Algorithm Training}
\nameeot~adopts a self-supervised training approach, learning the non-Markovian property inherent in offline dataset and the error information introduced due to the coupling between the state and extension in an end-to-end manner, as shown in Fig. \ref{eee}. For the variables in the proposed BF framework that lack closed-form expression, obtaining their true labels is challenging, making it impossible to directly learn the corresponding mapping functions. Therefore, we compute the joint error between the model's estimated posteriors $\mathbf{x}_{k|k}$ and $\mathbf{X}_{k|k}$ with the ground truth in the offline dataset, and then use the backpropagation method to update the model's weight parameters. Notably, the iterative formulas with compensation for the error terms are implemented through a Gaussian approximation, meaning that the closed-form of the proposed BF framework is differentiable. This enables the gradient of the error to propagate to the compensation terms we aim to learn, allowing the model to optimize the parameters. This learning method has been demonstrated as a viable paradigm in \cite{yan2024explainable}.

The stochastic gradient descent method is used for training   \nameeot. We use the minimum mean square error (MMSE) combined with additional L2 regularization as the loss function, which is consistent with the strategy commonly employed in sequential regression methods. For the offline dataset $\mathcal{D}$ given in Eq. \eqref{offline_dataset}, during the $t$-th training iteration, we denote the parameters in \nameeot~as $\bm{\varTheta}_t$. The prediction error for the input of the $j$-th sequence can be calculated as follows:
\begin{align}
    \ell_j (\bm{\varTheta}_t) & = \frac{1}{2K} \sum_{k=1}^{K} (\parallel \mathbf{x}^j_{k|k}(\mathbf{Z}^j_k; \bm{\varTheta}_t) - \bar{\mathbf{x}}_k^j \parallel^2 \nonumber \\
    & \ + \parallel \mathbf{X}^j_{k|k}(\mathbf{Z}^j_k; \bm{\varTheta}_t) - \overline{\mathbf{X}}_k^j \parallel^2) + \gamma \parallel \bm{\varTheta}_t \parallel^2.
\end{align}
where, $\gamma$ stands for a regularization coefficient applied to the L2 regularization term.

Both BF and RNNs operate in a similar recursive manner, processing data step-by-step. Likewise, \nameeot~shares the same mathematical formulation, making it compatible with backpropagation through time~\cite{werbos1990backpropagation}, which is widely used for training RNN-based DNNs.

\section{Experimental Study} \label{s_4}
In this section, we evaluate \nameeot~ using both simulated non-Markovian data and real-world data for joint state and extension estimation. The experiments demonstrate EOTNet’s effectiveness in addressing non-Markovian dynamics, nonlinearity, and the coupling of state and extension estimation. Comparative analysis with SOTA methods validates its superiority, and ablation studies verify the roles of the memory update and compensation blocks\footnote{The source code is available at: \url{https://github.com/Austin2wwzx/EOTNet}.}. The experiments are outlined as follows:
\begin{itemize}
    \item {\bf{EOT scenario simulation.}} This simulation validates the performance of \nameeot~in sequence filtering tasks with non-Markovian characteristics. Maneuvering target data, where non-linear trajectories induce temporal correlations equivalent to non-Markovian properties, is used for evaluation. \nameeot~in compared against several SOTA time-series models, including LSTM \cite{schmidhuber1997long}, Transformer \cite{vaswani2017attention}, and the MB EOT methods of \cite{feldmann2008tracking}, \cite{orguner2012variational}, and \cite{yang2019tracking}, in an EOT task with non-Markovian dynamics.
    \item {\bf{Extended landing aircraft tracking.}} This experiment evaluates the accuracy of \nameeot~in estimating trajectories and target shapes in realistic tracking scenarios with non-Markovian properties. Using observed aircraft landing trajectories, scatterers are generated based on the target’s position and extension, following the method in \cite{tuncer2021random}, to outline the aircraft’s shape. The comparative methods are the same as those used in the prior simulation, highlighting \nameeot’s performance in real-world tasks.
    \item {\bf{Ablation study.}} Finally, an ablation study evaluates the effectiveness of each DNN block of \nameeot. By masking specific gating units, we validate the model’s ability to compensate for mismatch errors arising from Markovian assumptions in joint state-extension estimation tasks.
\end{itemize}

Throughout the experiment, we use three metrics to quantify estimation errors: root mean square error (RMSE) for model accuracy, intersection over union (IoU) for overlap between predicted and ground truth bounding boxes, and Gaussian Wasserstein distance (GWD) for discrepancies between two Gaussian distributions. These metrics are detailed in the following:

{\em RMSE:} On a two-dimensional plane, the position RMSE of all state estimation posterior on test set is calculated as
\begin{align}
    \mathrm{RMSE} = \frac{1}{JK} \displaystyle\sum_{j=1}^{J} \displaystyle\sum_{k=1}^{K} \parallel \mathbf{x}^j_{1,2:k|k} - \bar{\mathbf{x}}^j_{1,2:k} \parallel^2,
\end{align}
where $J$ is the number of test cases, $K$ is the sequence length of each cases, $\mathbf{x}^j_{1,2:k|k}$, and $\bar{\mathbf{x}}^j_{1,2:k}$ stand for the estimated position and the ground truth position, respectively.

{\em IoU:}
The elliptical equation corresponding to the estimated and the ground truth extension is defined as
\begin{gather}
    \mathbf{ell}_{k|k} : \mathbf{x}_{1,2:k|k}^\mathrm{T}\mathbf{X}_{k|k}\mathbf{x}_{1,2:k|k}=1, \\
    \overline{\mathbf{ell}}_k : \bar{\mathbf{x}}_{1,2:k}^\mathrm{T}\overline{\mathbf{X}}_{k}\bar{\mathbf{x}}_{1,2:k}=1,
\end{gather}
where, $\mathbf{ell}_{k|k}$ and $\overline{\mathbf{ell}}_k$ stand for the target's estimated and ground truth extension at time $k$, respectively.
The IoU score of two ellipse in two-dimensional rectangular coordinate system at time $k$ is calculated as 
\begin{align}
    \mathrm{IoU}_k = \frac{\mathbf{ell}_{k|k} \cap \overline{\mathbf{ell}}_k}{\mathbf{ell}_{k|k} \cup \overline{\mathbf{ell}}_k},
\end{align}
where $\mathbf{ell}_{k|k} \cap \overline{\mathbf{ell}}_k$ is the area where the predicted ellipse and the ground truth ellipse overlap, $\mathbf{ell}_{k|k} \cup \overline{\mathbf{ell}}_k$ stands for the combined area covered by ellipses, with the overlapping area counted only once.

{\em GWD:}
The GWD compares two ellipses at time $k$:
\begin{align}
    \mathrm{GWD}_k & = \parallel \mathbf{x}_{1,2:k|k} - \bar{\mathbf{x}}_{1,2:k} \parallel ^ 2 \nonumber \\
    & + \mathrm{tr}[\mathbf{X}_{k|k} + \overline{\mathbf{X}}_{k} - 2 (\mathbf{X}_{k|k}^\frac{1}{2} \overline{\mathbf{X}}_{k} \mathbf{X}_{k|k}^\frac{1}{2})^\frac{1}{2}].
\end{align}

\subsection{EOT Scenario Simulation} \label{simulation}

In the simulation, we compare our model with two advanced time-series deep learning models, LSTM and Transformer, both utilizing step-by-step generation to align with real-world filtering causality. Additionally, we evaluate our method against three widely used EOT algorithms: Feldmann’s IMM-RMM \cite{feldmann2008tracking}, Orguner’s VB-RMM \cite{orguner2012variational}, and Yang’s MEM-EKF \cite{yang2019tracking}. IMM-RMM integrates interacting multiple models with RMM for reduced response latency in maneuvering target scenarios. VB-RMM applies variational Bayesian inference for more accurate posterior estimates, albeit at higher computational cost. MEM-EKF introduces a novel measurement approach, decoupling orientation estimation from the major and minor axes of elliptical targets for improved extension parameter estimation. All these MB approaches rely on the first-order Markov assumption, consistent with the modeling in Eqs. \eqref{state_pred}–\eqref{measurement}.

{\bf{Simulation setup.}} The  scenario encompasses a target object with an unknown extension, with no false alarms or missed detections in sensor measurements. The tracked target maintains a constant size in  its shape (i.e., the lengths of the major and minor axes of the ellipse), while its orientation changes in time.
 
The target trajectories are generated similarly to \cite{liu2020deepmtt}, where state sequences with extensions are created within a specified area. Each sequence comprises the joint ground truth of the state and extension, along with noisy position measurements at each time step. The sequence length is set as $K=140s$, with a $1s$ interval between consecutive frames. The state is generated using two kinematic models—constant velocity (CV) and coordinate turn (CT)—with process noise, randomly switching during sequence generation to introduce randomness and non-Markovian characteristics. The target’s major and minor axes remain fixed at $10m$ and $2m$, respectively, and it moves at a constant speed of $10m/s$. Measurements are uniformly distributed on the elliptical extension, with the number of measurements per time step drawn from a Poisson distribution with a mean of $20$, as in \cite{lan2014tracking}.

The process and measurement noise are modeled as $\mathbf{w}_k \!\! \sim \!\! \mathcal{N}(\mathbf{0}, \sigma_w^2 \mathbf{I}_4)$ and $\mathbf{v}_k \!\! \sim \!\! \mathcal{N}(\mathbf{0}, \sigma_v^2 \mathbf{I}_2)$, respectively. Four datasets are generated with noise levels \{$\sigma_w, \sigma_v$\} \!\! = \!\! \{0.4,0.6\}, \{0.6,0.8\}, \{0.8,1\} and \{1,1.2\}, each containing 600 cases (480 for training and 120 for testing). For DB models, we use 5-fold cross-validation on the training sets to optimize model weights for better generalization.

Considering the difficulty of accurately modeling the motion of the target, the nominal state evolution model of \nameeot~and others MB methods is set as a CV model, namely, the nominal state evolution function is
\begin{gather}
    f_k(\mathbf{x}_{k-1}) = \begin{bmatrix}
        1 & 0 & \Delta t & 0 \\
        0 & 1 & 0 & \Delta t \\
        0 & 0 & 1 & 0 \\
        0 & 0 & 0 & 1
        \end{bmatrix} \mathbf{x}_{k-1}, \label{fffff}
\end{gather}
with the sampling interval $\Delta t=1$. This experiment has a higher dimension compared to the one-dimensional time series, so we appropriately increase the learnable parameter to set both the number of hidden layer nodes and the dimension of the memory vector in \nameeot~to 64.
A simple linear model is used to model the observations for \nameeot~and others MB methods, i.e., the nominal state measurement function is
\begin{gather}
    h_k(\mathbf{x}_{k}) = \begin{bmatrix}
        1 & 0 & 0 & 0 \\
        0 & 1 & 0 & 0 
        \end{bmatrix} \mathbf{x}_{k} .\label{hhhhh}
\end{gather}

The DB comparison methods are as follows: 1) Transformer: A single-layer autoregressive Transformer block with 8 self-attention heads and an embedded feature dimension of 128, using the previous estimated state and current measurement as inputs. 2) LSTM: A single-layer LSTM with the same input-output structure as the Transformer, and 64 hidden nodes.

The parameter settings for MB methods are as follows: For both RMMs, $v$ is set to $7$. The base filter for IMM-RMM is EKF, with a model set including one constant acceleration (CA) model, one CV model, and two CT models with turn rates of $-3^\mathrm{o}/s$ and $3^\mathrm{o}/s$, respectively. The VB-RMM follows \cite{orguner2012variational}, with 3 variational iterations. For MEM-EKF, the prior $k$ of the shape variables is speciﬁed by the covariance matrix $\mathbf{C}_0^p=\textrm{diag}[1, 1]$. The process noise covariance is set to $\mathbf{C}_p^w=\textrm{diag}[0.1, 0.1]$, and the transition matrix is $\mathbf{A}_k^p=\mathbf{I}_3$. 

{\bf{Experimental results and analysis.}} We compared the predictive performance of trained DB and MB methods across four noise levels. The mean RMSE values for position are shown in Tables \ref{pos_mse}, where ``Trans.'', ``IMM.'', ``VB.'', and ``MEM.'' represent Transformer, IMM-RMM, VB-RMM, and MEM-EKF, respectively. In position estimation, \nameeot~uses a priori state transition functions, effectively constraining the feature search space compared to DB methods, leading to better accuracy. Additionally, its memory mechanism compensates for model mismatch errors from the first-order Markov assumption, resulting in lower estimation errors than MB methods.

\begin{figure}[t]
    \centering
    \includegraphics[width=1\linewidth]{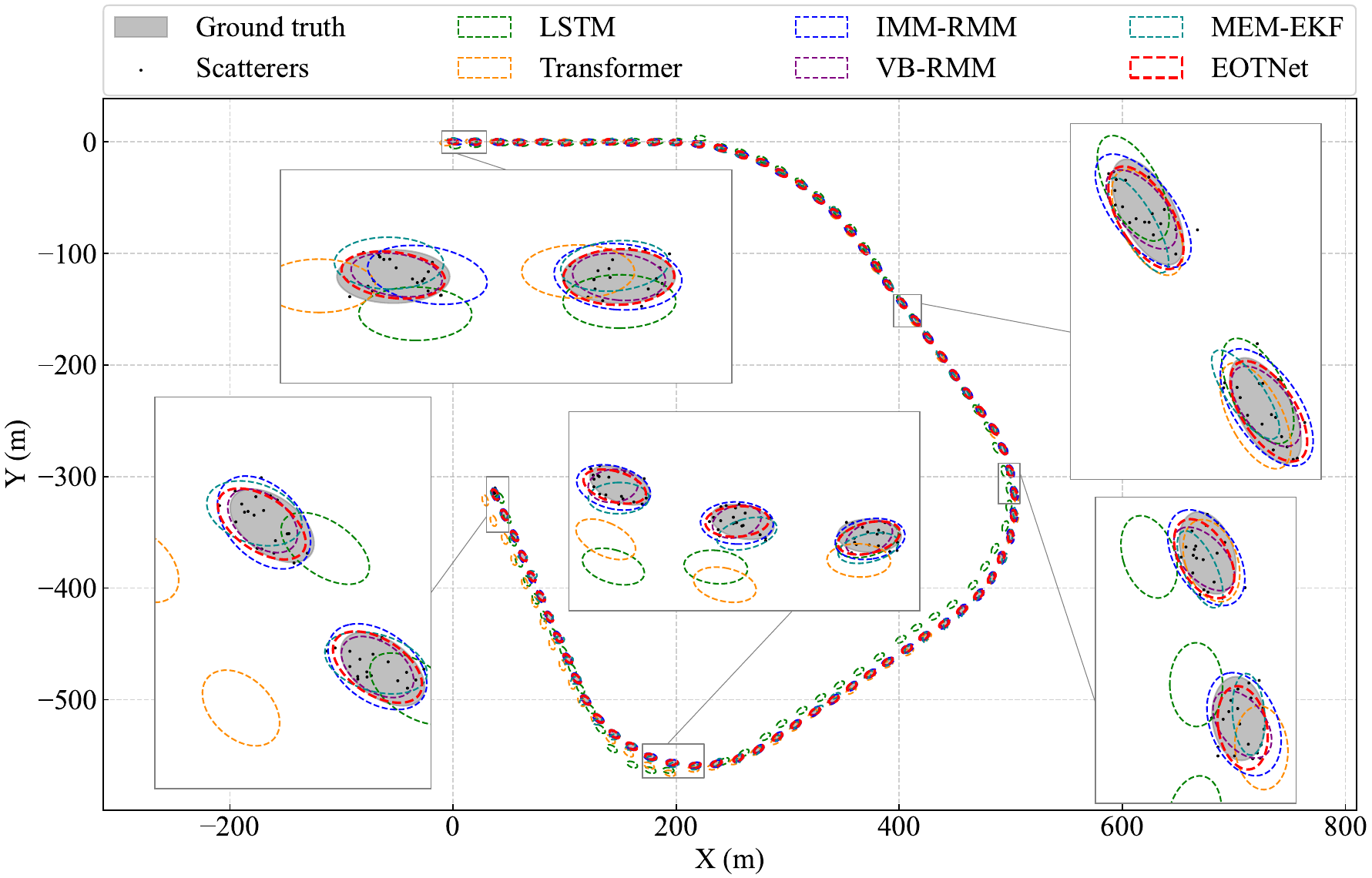}
    \caption{The measurements, trajectory, and estimation results of a single example run on test set with a noise level $\sigma_w$=0.4 and $\sigma_v$=0.6.\protect}
    \label{eot_result}
\end{figure}

\begin{figure}
    \centering
    \includegraphics[width=1\linewidth]{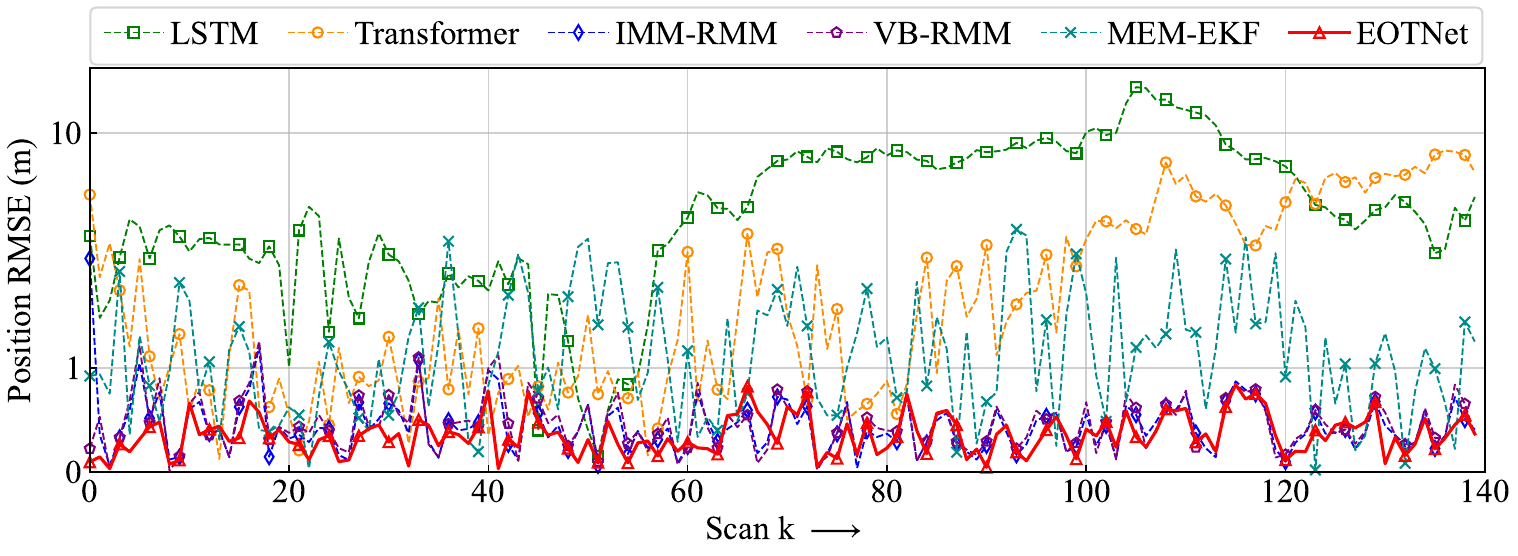}
    \vspace{-2em}
    \caption{Position estimation RMSE of the test sample.\protect}
    \label{pos_RMSE}
\end{figure}

\begin{figure}
    \centering
    \includegraphics[width=1\linewidth]{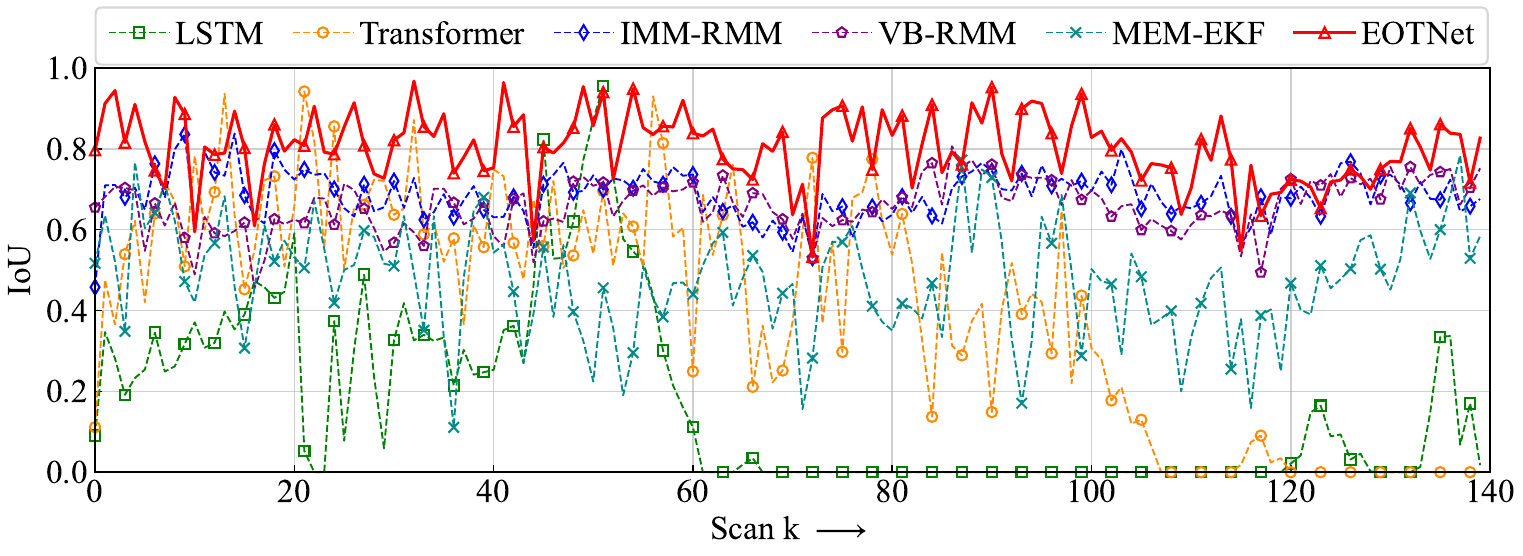}
    \vspace{-2em}
    \caption{Extension accuracy based on the IoU of the test sample.\protect}
    \label{iou}
\end{figure}

\begin{figure}
    \centering
    \includegraphics[width=1\linewidth]{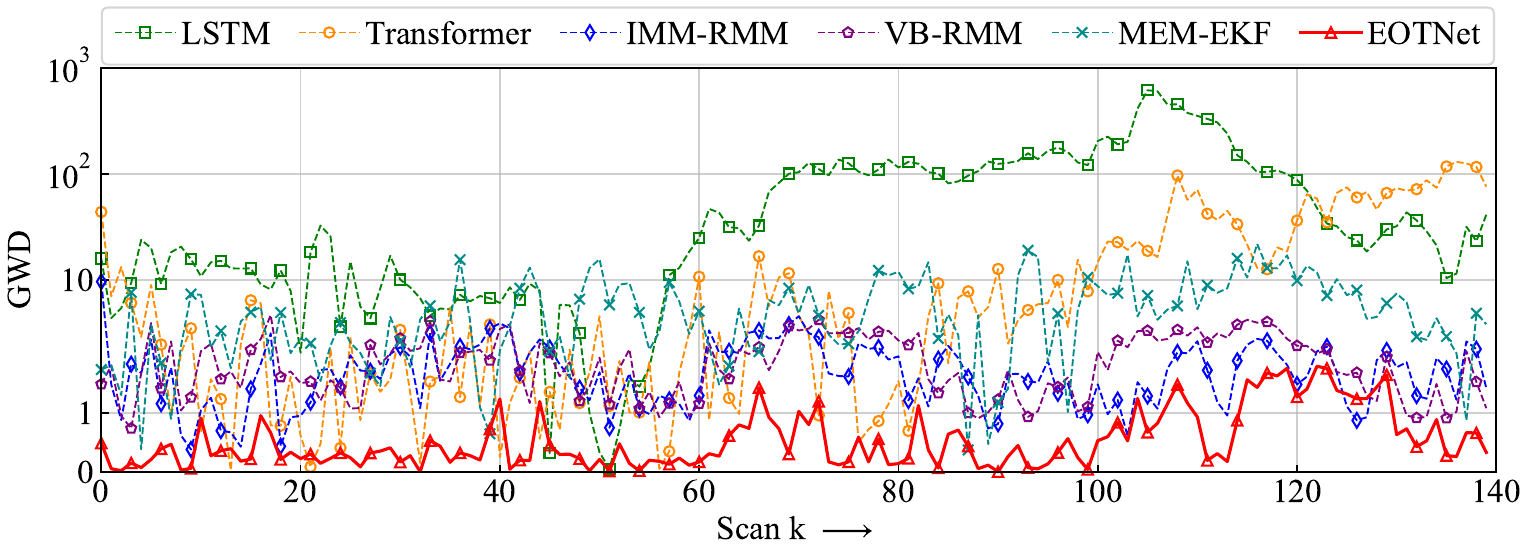}
    \vspace{-2em}
    \caption{Extension estimation error based on the GWD of the test sample.\protect}
    \label{gwd}
\end{figure}

The IoU values and GWD for the target’s extension estimation are shown in Tables \ref{iou_result} and \ref{gwd_result}. Due to higher position estimation errors, LSTM and Transformer exhibit lower IoU values and higher GWD, reflecting poor shape estimation. In contrast, the memory mechanism of \nameeot~captures the coupling between state and extension, leading to more accurate shape estimation than the MB methods.
\vspace{-0.4em}
\begin{table}[H]
    \caption{Mean position RMSE ($m$) on different test sets.}
    \label{pos_mse}
    \centering
    \vspace{-0.4em}
    \resizebox{\columnwidth}{!}{
    \begin{tabular}{c|cccccc}
    \toprule
    $\sigma_w$, $\sigma_v$ & LSTM & Trans. & IMM. & VB. & MEM. & \nameeot \\
    \midrule
    0.4, 0.6 & 7.012 & 2.704 & 0.508 & 0.522 & 1.227 & \bf{0.398} \\
    0.6, 0.8 & 7.621 & 3.628 & 0.535 & 0.554 & 1.267 & \bf{0.429} \\
    0.8, 1 & 7.548 & 3.883 & 0.610 & 0.636 & 1.368 & \bf{0.501} \\
    1, 1.2 & 7.441 & 3.804 & 0.737 & 0.780 & 1.531 & \bf{0.623} \\
    \bottomrule
    \end{tabular}}
\end{table}
\vspace{-1.2em}
\begin{table}[H]
    \caption{Mean values of IoU on different test sets.}
    \label{iou_result}
    \centering
    \vspace{-0.4em}
    \resizebox{\columnwidth}{!}{
    \begin{tabular}{c|cccccc}
    \toprule
    $\sigma_w$, $\sigma_v$ & LSTM & Trans. & IMM. & VB. & MEM. & \nameeot \\
    \midrule
    0.4, 0.6 & 0.174 & 0.473 & 0.697 & 0.667 & 0.500 & \bf{0.815} \\
    0.6, 0.8 & 0.087 & 0.381 & 0.654 & 0.672 & 0.512 & \bf{0.790} \\
    0.8, 1 & 0.126 & 0.414 & 0.577 & 0.681 & 0.481 & \bf{0.764} \\
    1, 1.2 & 0.168 & 0.387 & 0.475 & 0.666 & 0.453 & \bf{0.725} \\
    \bottomrule
    \end{tabular}}
\end{table}
\vspace{-1.2em}
\begin{table}[H]
    \caption{Mean GWD ($m$) on different test sets.}
    \label{gwd_result}
    \centering
    \vspace{-0.4em}
    \resizebox{\columnwidth}{!}{
    \begin{tabular}{c|cccccc}
    \toprule
    $\sigma_w$, $\sigma_v$ & LSTM & Trans. & IMM. & VB. & MEM. & \nameeot \\
    \midrule
    0.4, 0.6 & 98.420 & 14.637 & 1.633 & 1.887 & 4.949 & \bf{0.483} \\
    0.6, 0.8 & 116.431 & 26.343 & 2.067 & 1.930 & 5.187 & \bf{0.683} \\
    0.8, 1 & 114.085 & 30.178 & 3.268 & 1.944 & 5.853 & \bf{0.792} \\
    1, 1.2 & 110.811 & 28.959 & 5.958 & 2.177 & 7.091 & \bf{0.963} \\
    \bottomrule
    \end{tabular}}
\end{table}
\vspace{-0.5em}
Fig. \ref{eot_result} shows the measurements, trajectory, and estimation results from a single run on the test set with noise levels $\sigma_w$=0.4 and $\sigma_v$=0.6. Table \ref{average_peak_score} presents the average performance and extreme cases of various methods, based on Pos. RMSE, IoU, and GWD metrics for the sequence in Fig. \ref{eot_result}. The peak IoU value represents the lowest case, while for the other metrics, it indicates the highest error case. Figs. \ref{pos_RMSE} - \ref{gwd} show the prediction results at different time steps.

The figures and table show that \nameeot~outperforms other methods in EOT tasks with non-Markovian characteristics, achieving superior average and peak performance. Purely DB methods struggle with position estimation due to the lack of kinematic modeling, leading to significantly higher errors and a larger spatial discrepancy. The MB method, including VB-RMM and MEM-EKF, rely on a naive CV model, provides superior state and extension estimation accuracy but struggle in non-Markov conditions. IMM-RMM performs well in slow maneuvering scenarios; however, it relies on accurate prior model switching probabilities, and its performance degrades when these probabilities are mismatched. Overall, \nameeot~demonstrates more stable and accurate estimation than all MB methods, which assume first-order Markov processes.

\begin{table}
    \belowrulesep=0pt
    \aboverulesep=0pt
    \caption{Average and peak estimation performance of the single example.}
    \label{average_peak_score}
    \vspace{-2mm}
    \begin{tabularx}{\linewidth}{>{\raggedright\arraybackslash}X |>{\centering\arraybackslash}p{2.1cm} >{\centering\arraybackslash}X >{\centering\arraybackslash}X}
    \toprule
    \multirow{2}{*}{Methods} & \multicolumn{3}{c}{Average value} \\
    \cmidrule(lr){2-4}
     & Pos. RMSE (m) & IoU & GWD \\
    \midrule
    \raggedright LSTM & 5.0753 & 0.1711 & 79.0308 \\
    \raggedright Transformer & 2.2788 & 0.4082 & 18.2918 \\
    \raggedright IMM-RMM & 0.4433 & 0.6850 & 1.8312 \\
    \raggedright VB-RMM & 0.4638 & 0.6587 & 2.0779 \\
    \raggedright MEM-EKF & 1.1407 & 0.4787 & 5.9066 \\
    \raggedright \nameeot & \bf{0.3585} & \bf{0.8023} & \bf{0.5396} \\
    \midrule
     & \multicolumn{3}{c}{Peak value} \\
    \cmidrule(lr){2-4}
     & Pos. RMSE (m) $\downarrow$ & IoU $\uparrow$ & GWD $\downarrow$ \\
    \midrule
    \raggedright LSTM & 17.5619 & 0 & 616.8412 \\
    \raggedright Transformer & 8.0656 & 0 & 130.2596 \\
    \raggedright IMM-RMM & 2.1393 & 0.4577 & 9.6859 \\
    \raggedright VB-RMM & 1.2572 & 0.4551 & 4.6809 \\
    \raggedright MEM-EKF & 3.0606 & 0.1109 & 21.9672 \\
    \raggedright \nameeot & \bf{0.8417} & \bf{0.5335} & \bf{1.7974} \\
    \bottomrule
    \end{tabularx}
\end{table}


\begin{figure}
    \centering
    \includegraphics[width=1\linewidth]{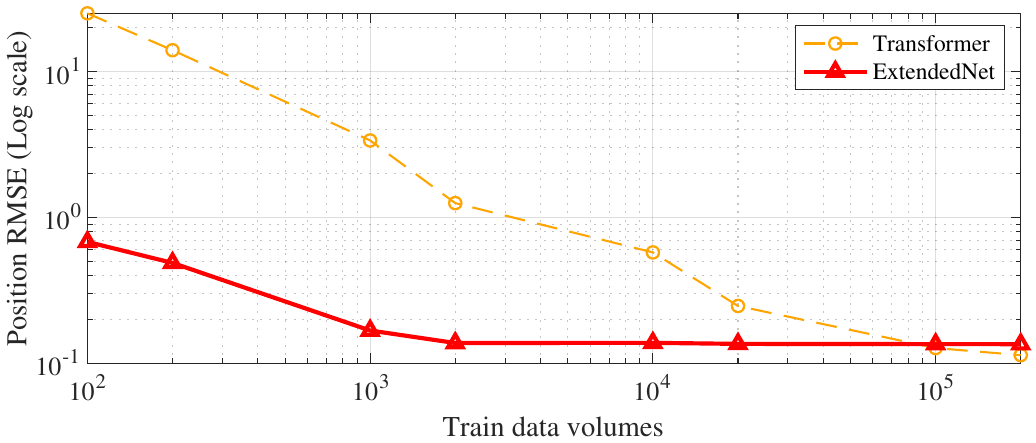}
    \caption{Position estimation RMSE under different train data volumes.\protect}
    \label{asd}
\end{figure}

To compare the data dependency of DB methods such as Transformer and MB learning methods such as \nameeot with optimal prediction accuracy, we performed experiments with varying volumes of training data in a simulation scenario. The parameter settings were consistent with previous experiments, under noise levels $\sigma_w$ = 0.4 and $\sigma_v$ = 0.6. Datasets with 100, 200, 1,000, 2,000, 10,000, 20,000, 100,000, and 200,000 trajectories were generated, along with a randomized test set of 2,000 trajectories. We trained Transformer and \nameeot separately with different training volumes and compared their performance on the test set. The logarithm of the position estimation RMSE is shown in Fig. \ref{asd}.

The experimental results show that \nameeot~is far less dependent on training data volumes than purely DB methods. Even with limited data, it maintains stable estimation accuracy due to its integration of prior knowledge within the BF framework, which reduces the parameter search space. In contrast, Transformer, lacking physical priors, struggles with sparse data, leading to underfitting and poor state estimation. While its accuracy improves with more data, \nameeot consistently outperforms DB methods, making it particularly advantageous in data-limited scenarios. As the volume of data increases, the performance of the Transformer outperforms that of EOTNet when the dataset contains 200,000 train cases. However, acquiring such large volumes of data is often unrealistic in practical tasks.

\subsection{Extended Landing Aircraft Tracking}
We used the Northern California TRACON dataset, a publicly available dataset representing realistic flight operations, to further evaluate each methods. It captures approach trajectories of aircraft influenced by control tower instructions and turbulence, resulting in non-Markovian characteristics that make joint state and extension estimation challenging. Using the same comparative methods from Section \ref{simulation}, we assess EOTNet's ability to address these challenges by leveraging partial prior knowledge.

{\bf{Experimental setup.}} We extracted four months of aircraft landing trajectories for a single runway. To standardize temporal intervals, Laplacian interpolation fixed the frame interval to $\Delta t$ = 4. Trajectories were then segmented into 200 time steps, truncating longer sequences and excluding shorter ones, resulting in 260 sequences. Of these, 208 were used for training, and 52 for evaluation. Following Section \ref{simulation}, 5-fold cross-validation was applied to optimize model weights for generalization.

To ensure data authenticity, we used the pre-processing method from \cite{tuncer2021random}. The target is modeled as a civil aircraft with a length of 73.9 meters and width of 64.8 meters, with scatterers uniformly distributed over its surface. We introduced four noise levels, i.e., 25m, 50m, 100m, and 150m, to reflect varying radar accuracy in real-world scenarios and test the generalizability of different methods. The state transition and measurement processes for the MB methods and \nameeot~follow the procedures in Eqs. \eqref{fffff} and \eqref{hhhhh}.

In real-world scenarios, the joint state and extension estimation exhibit long-term contextual dependencies and non-Markovian effects. To capture these complexities, the feature dimension for \nameeot~is set to 96. For the Transformer, the embedding dimension is 128 with 16 attention heads to capture rich feature relationships. The LSTM uses 128 hidden nodes, matching the Transformer’s embedding dimension for consistent feature representation.

The parameter settings for MB methods are as follows: $v$ is set to $3$ in both RMMs. The settings of VB-RMM references to \cite{orguner2012variational}, its variational iterations are set to 10. The prior $k$ of the shape variables for MEM-EKF is speciﬁed by the covariance matrix $\mathbf{C}_0^p=\textrm{diag}[10, 10]$. All other parameters remain consistent with those used in the simulation scenarios describe before.

{\bf{Experimental results and analysis.}}
The mean position RMSE, IoU, and GWD for each method on the test set are shown in Tables \ref{la_pos_rmse}, \ref{la_iou_result}, and \ref{la_gwd_result}. Compared to DB methods, \nameeot improves position estimation accuracy by integrating prior knowledge through the Bayesian filter, significantly reducing IoU and GWD. Compared to MB methods, \nameeot better extracts state and extension modes from offline data and compensates for model mismatch errors, resulting in the lowest estimation errors across all metrics.

\begin{figure}[H]
    \centering
    \includegraphics[width=1\linewidth]{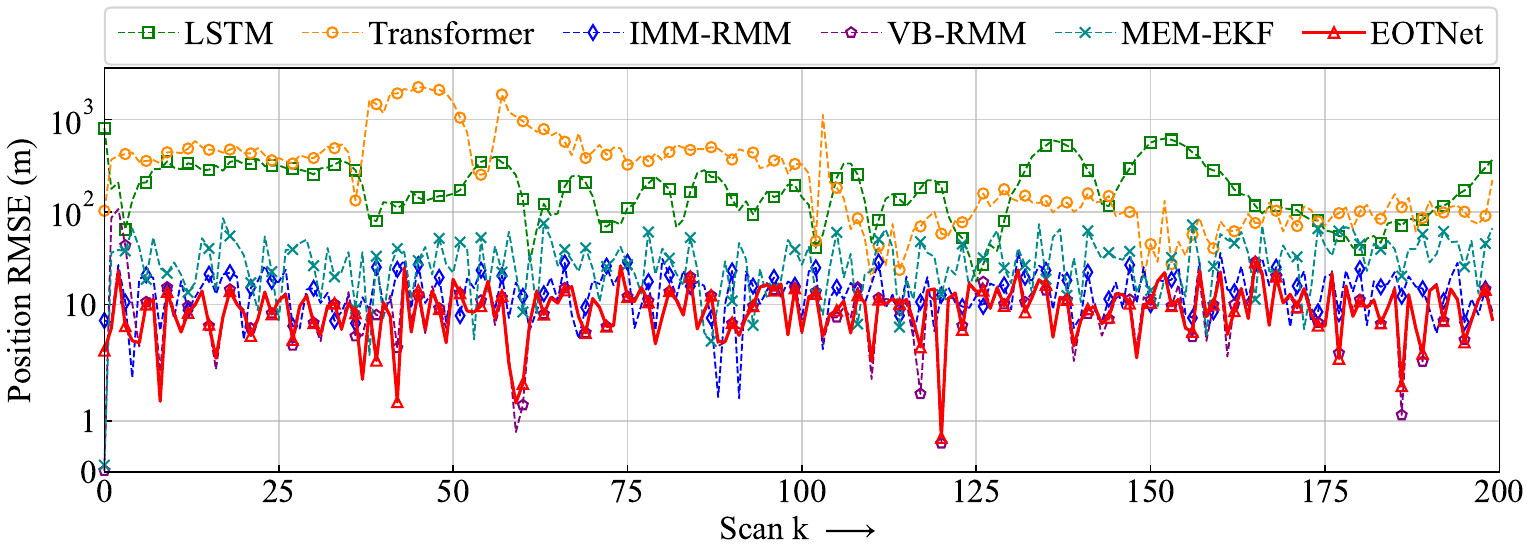}
    \caption{Position estimation RMSE of the test sample.\protect}
    \label{la_pos_RMSE}
\end{figure}

\begin{figure}[H]
    \centering
    \includegraphics[width=1\linewidth]{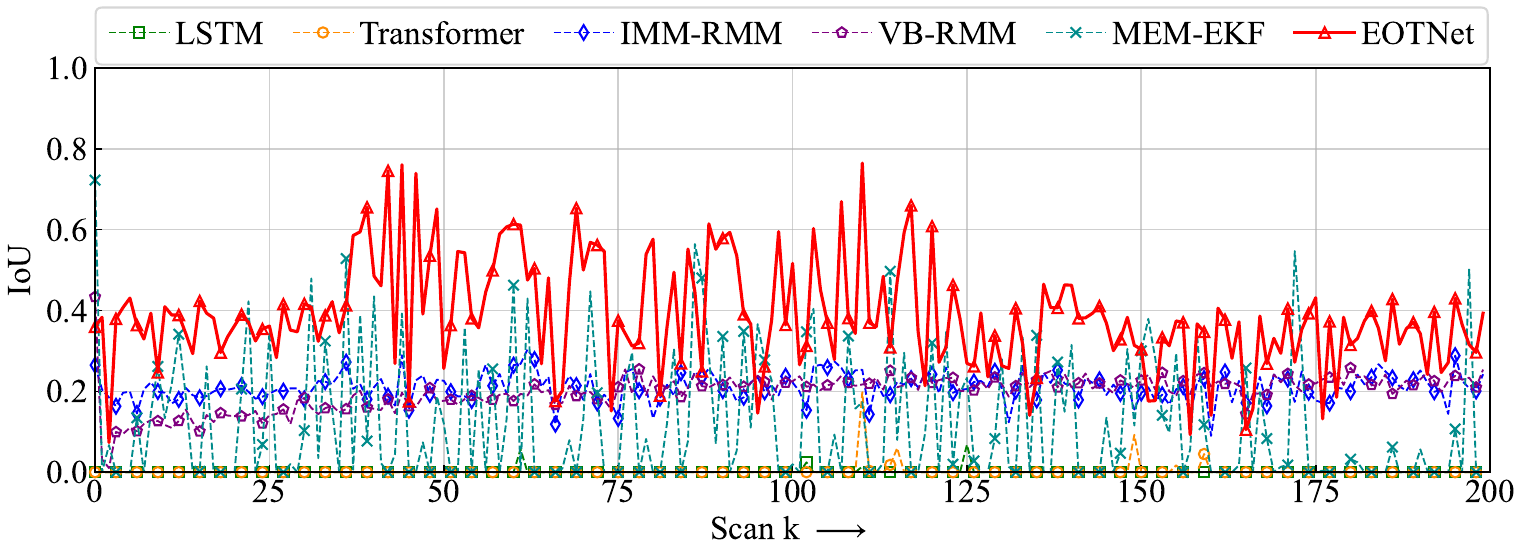}
    \vspace{-1.5em}
    \caption{Extension accuracy based on the IoU of the test sample.\protect}
    \label{la_iou}
\end{figure}

\begin{figure}[H]
    \centering
    \includegraphics[width=1\linewidth]{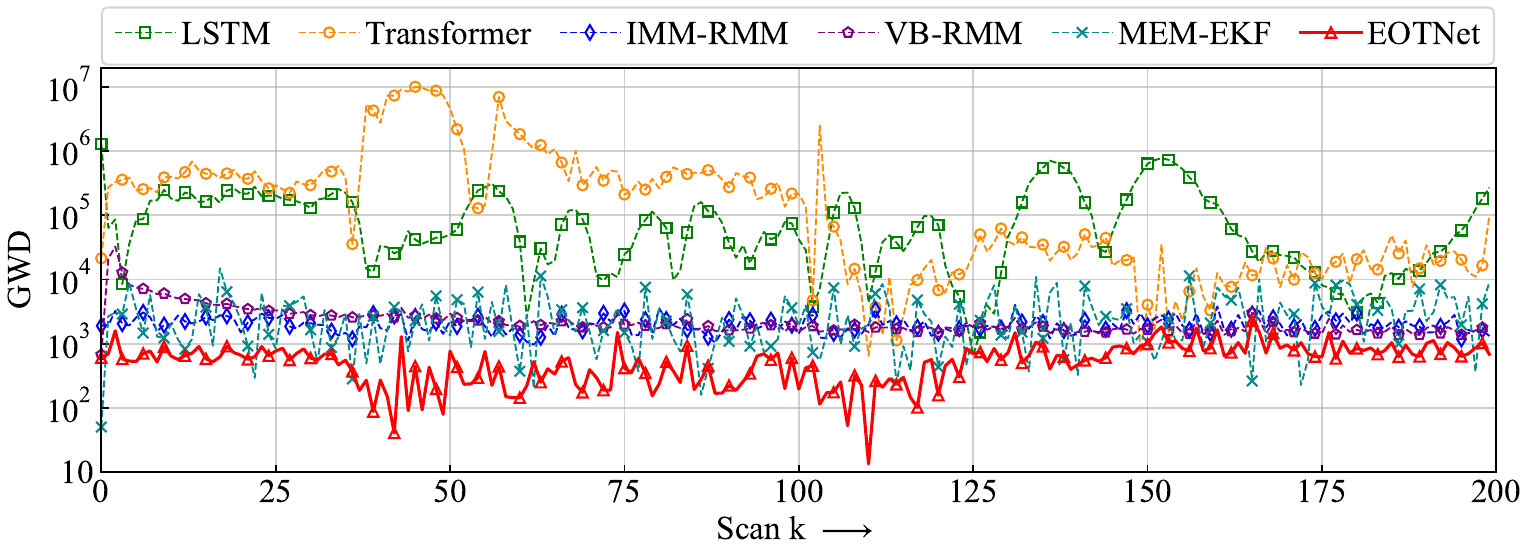}
    \vspace{-1.5em}
    \caption{Extension estimation error based on the GWD of the test sample.\protect}
    \label{la_gwd}
\end{figure}

Figs. \ref{la_pos_RMSE} - \ref{la_gwd} show the prediction results at different time steps. Despite some fluctuations, \nameeot~consistently outperforms other methods in prediction accuracy, demonstrating its strong adaptability to complex non-Markovian characteristics and its ability to capture the intricate relationships between state and extension for more accurate estimations in real-world EOT tasks.
\begin{table}[H]
    \caption{Mean position RMSE ($m$) on the Landing aircraft test sets with different noise levels.}
    \label{la_pos_rmse}
    \centering
    \resizebox{\columnwidth}{!}{
    \begin{tabular}{c|cccccc}
    \toprule
    $\sigma_w$ & LSTM & Trans. & IMM. & VB. & MEM. & \nameeot \\
    \midrule
    25 & 481.314 & 353.953 & 9.496 & 7.204 & 20.667 & \bf{6.682} \\
    50 & 422.861 & 366.301 & 16.673 & 16.161 & 34.298 & \bf{11.613} \\
    100 & 468.014 & 314.225 & 27.122 & 27.594 & 60.619 & \bf{22.533} \\ 
    150 & 478.324 & 405.037 & 39.765 & 38.667 & 82.607 & \bf{32.590} \\   
    \bottomrule
    \end{tabular}}
\end{table}
\vspace{-16pt}
\begin{table}[H]
    \caption{Mean values of IoU ($\times 10^{-2}$) on the Landing aircraft test sets with different noise levels.}
    \label{la_iou_result}
    \centering
    \resizebox{\columnwidth}{!}{
    \begin{tabular}{c|cccccc}
    \toprule
    $\sigma_w$ & LSTM & Trans. & IMM. & VB. & MEM. & \nameeot \\
    \midrule
    25 & 0.359 & 0.995 & 49.134 & 53.881 & 25.678 & \bf{53.201} \\
    50 & 0.339 & 1.016 & 20.825 & 19.974 & 11.984 & \bf{37.397} \\
    100 & 0.319 & 1.464 & 5.959 & 5.479 & 5.241 & \bf{21.827} \\
    150 & 0.248 & 0.929 & 2.713 & 1.148 & 3.290 & \bf{13.413} \\    
    \bottomrule
    \end{tabular}}
\end{table}
\vspace{-16pt}
\begin{table}[H]
    \caption{Mean GWD ($\times 10^{3}m$) on the Landing aircraft test sets with different noise levels.}
    \label{la_gwd_result}
    \centering
    \resizebox{\columnwidth}{!}{
    \begin{tabular}{c|cccccc}
    \toprule
    $\sigma_w$ & LSTM & Trans. & IMM. & VB. & MEM. & \nameeot \\
    \midrule
    25 & 463.392 & 250.591 & \bf{0.316} & 0.274 & 1.055 & 0.319 \\
    50 & 357.687 & 268.383 & 2.050 & 2.858 & 2.885 & \bf{0.632} \\
    100 & 438.138 & 197.504 & 12.709 & 14.681 & 7.588 & \bf{1.301} \\
    150 & 457.657 & 328.147 & 33.716 & 89.860 & 13.861 & \bf{2.411} \\ 
    \bottomrule
    \end{tabular}}
\end{table}
\vspace{-16pt}
\vspace{-16pt}
\vspace{-16pt}
\vspace{-1em}
\subsection{Ablation Study}
To validate the mismatch compensation mechanism in \nameeot~under the non-Markov properties, we conducted ablation experiments on the simulated dataset from Section \ref{simulation}, selectively masking the JEB, JUB, and MUB components.

{\bf{Experimental setup.}} 
We conducted experiments on the dataset with noise levels, {$\sigma_w$=1, $\sigma_v$=1.2}, using the same \nameeot parameters as in Section \ref{simulation}. The methods included: 1) the full \nameeot as the control group; 2) (w/o) JEB, masking JEB to assess compensation for evolution mismatch errors; 3) (w/o) JUB, masking JUB to evaluate compensation for measurement mismatch errors; and 4) (w/o) MUB, masking MUB by removing the LSTM Cell and using only the previous state and extension as inputs for JEB and JUB.

\begin{figure}[H]
    \centering
    \includegraphics[width=1\linewidth]{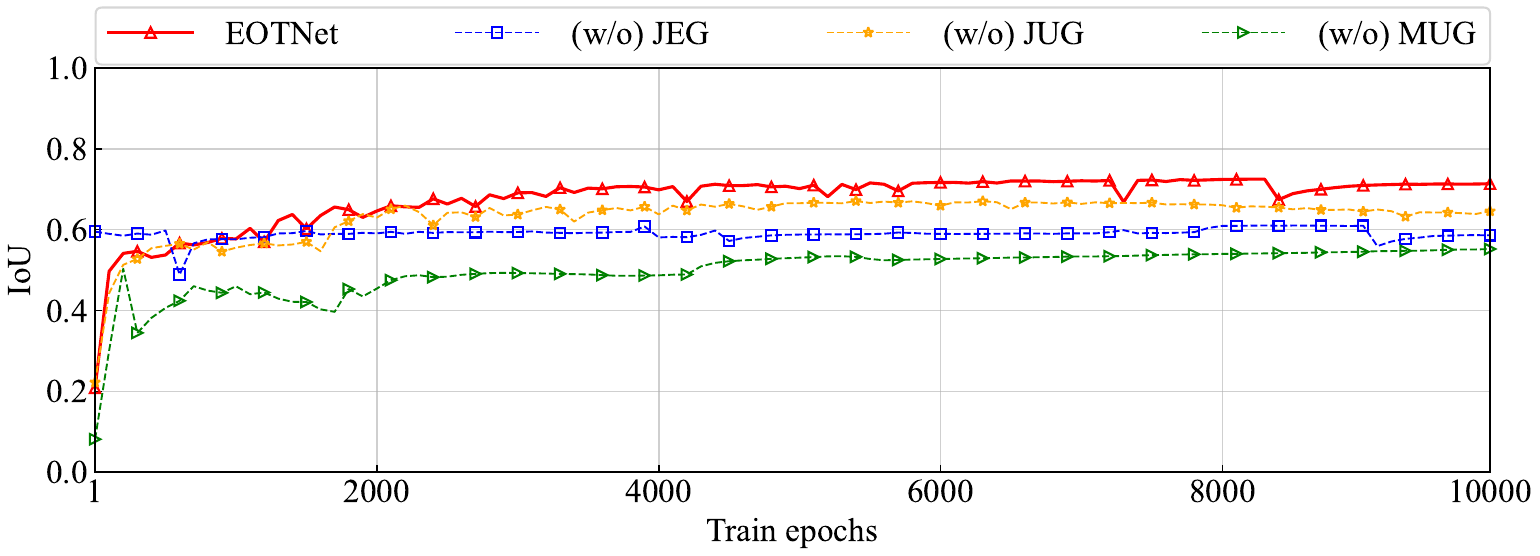}
    \caption{IoU score on the test set during training process. \protect}
    \label{ac_iou}
\end{figure}

\begin{figure}[H]
    \centering
    \includegraphics[width=1\linewidth]{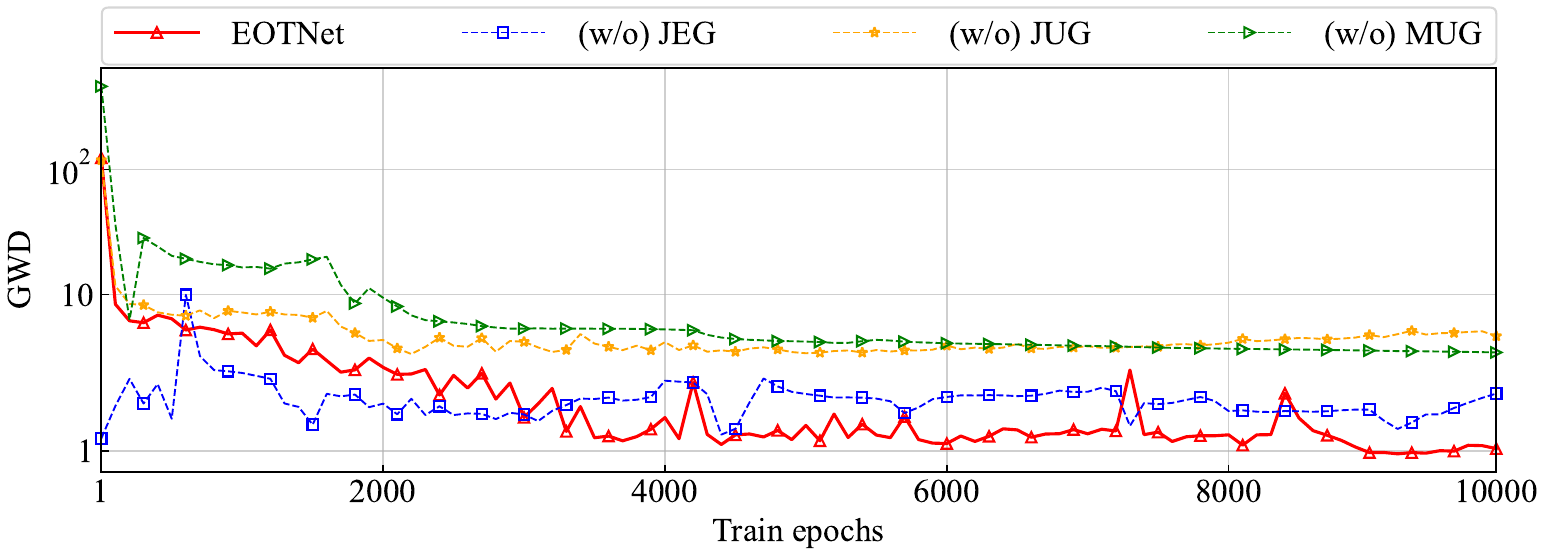}
    \caption{GWD on the test set during training process. \protect}
    \label{ac_gwd}
\end{figure}

{\bf{Experimental results and analysis.}} Figs. \ref{ac_iou} and \ref{ac_gwd} illustrate the IoU score and GWD of \nameeot~and its three variant models throughout the training process. The complete \nameeot outperforms other models, highlighting the effectiveness of the three NN blocks in improving estimation accuracy. Masking JEB prevents proper compensation for evolution mismatch, leading to errors in prior estimates during updates. Masking JUB, which relies on basic linear measurements, limits the ability to accurately characterize the target’s extension, quickly reaching an accuracy ceiling. Without MUB, the loss of historical state and extension summarized by memory reduces \nameeot to a first-order Markov model, significantly impairing estimation accuracy during both evolution and updates.
\vspace{-1em}
\vspace{-2em}  %
\vspace{-5mm}
\begin{table}[H]
    \caption{Ablation study results with different methods on the test set with $\sigma_w$=1, $\sigma_v$=1.2.}
    \label{ab}
    \vspace{-2mm}
    \centering
    \begin{tabular}{c|cccc}
    \toprule
    Metrics & (w/o) JEB & (w/o) JUB & (w/o) MUB & \nameeot \\
    \midrule
    Pos. RMSE $\downarrow$ & 0.649 & 0.723 & 0.720 & \bf{0.622} \\
    IoU $\uparrow$ & 1.185 & 3.340 & 3.441 & \bf{0.963} \\
    GWD $\downarrow$ & 0.610 & 0.671 & 0.552 & \bf{0.725} \\
    \bottomrule
    \end{tabular}
\end{table}
Table \ref{ab} summarizes the optimal prediction performance of \nameeot~and its three variants, with checkpoints selected based on the best joint estimation accuracy on the test set. The complete \nameeot achieves superior accuracy in both position and extension estimation, highlighting the effectiveness of its gating compensation units over traditional RMM-based EOT methods. The MUB captures the coupling effects and non-Markovian dependencies, which are compensated by JEB during evolution. JUB further corrects model mismatch errors during updates, enhancing joint state-extension estimation accuracy.

\section{Conclusion} \label{s_5}
In this paper, we propose \nameeot, a deep learning augmented BF framework for joint state and extension estimation in non-Markovian tasks. Using Gaussian approximation, the BF framework derives an analytical expression and constructs a closed-form iterative process. \nameeot integrates this framework with a neural network, which captures model mismatch from offline data and incorporates it as deep memory. This approach enables efficient recursive estimation while reducing errors from model mismatch and state-extension coupling. Comparative experiments on simulated and real-world datasets show that \nameeot outperforms MB and DB methods, with lower reliance on extensive training data than DB approaches.


\appendix

\section*{Proof of Theorem~\ref{T_2}}
\label{app_C}
The mean of the state prediction is calculated as follows:
\begin{align}
    \!\!\!\mathbf{x}_{k|k-1} &= \mathbb{E}_{\mathbf{x}_k|\mathbf{z}_{1:k-1},\mathcal{D}}[\mathbf{x}_k] \nonumber \\
    &= \iiint \mathbf{x}_kp(\mathbf{x}_k,\mathbf{X}_k,\mathbf{c}_k|\mathbf{z}_{1:k-1},\mathcal{D})\mathrm{d}\mathbf{x}_k\mathrm{d}\mathbf{X}_k\mathrm{d}\mathbf{c}_k. \label{x_pppred}
\end{align}

Plugging Eqs. \eqref{asdasd} - \eqref{delta_f}, \eqref{ex_pred_mis} - \eqref{dasdas} into Eq. \eqref{x_pppred} we have
\begin{gather}
    \mathbf{x}_{k|k-1} = \int f_k(\mathbf{x}_{k-1}) P_{k-1}^x \mathrm{d}\mathbf{x}_{k-1} + \hat{\bm{\Delta}}_k^f, \label{state_prediction_temp}
\end{gather}
where $P_{k-1}^x = \mathcal{N}(\mathbf{x}_{k-1};\mathbf{x}_{k-1|k-1}, \mathbf{P}_{k-1|k-1})$.

However, the $f_k(\cdot)$ in the above expression may represent a nonlinear state transition process, making this integral challenging to compute. We employed a first-order Taylor's formula for linearization, similar to that used in the Extended Kalman Filter (EKF)\cite{durbin2012time}. Employing it into Eq. \eqref{state_prediction_temp}, then we have Eq. \eqref{state_pred_final}.

The covariance of the state prediction is calculated as follows:
\begin{gather}
    \mathbf{P}_{k|k-1} = \mathbb{E}_{\mathbf{x}_k|\mathbf{z}_{1:k-1},\mathcal{D}}[(\mathbf{x}_k - \mathbf{x}_{k|k-1})(\mathbf{x}_k - \mathbf{x}_{k|k-1})^\mathrm{T}] \nonumber \\
    \!\!\!\!\!\! = \!\!\! \int \!\!\!\! \int \!\!\!\! \int \!\!\!\! \int \!\!\!\! \int \!\!\!\! \int \!\!\!\! \int \!\! \mathbf{M}_k^f P_k^2\mathrm{d}\mathbf{x}_{k-1}\mathrm{d}\mathbf{X}_{k-1}\mathrm{d}\mathbf{c}_{k-1} \mathrm{d}\mathbf{X}_k \mathrm{d}\mathbf{c}_k \mathrm{d}\bm{\Delta}_k^f \mathrm{d}\bm{\Delta}_k^\phi, \label{X_pppred}
\end{gather}
with $\mathbf{M}_k^f = (f_k(\mathbf{x}_{k-1})+\bm{\Delta}_k^f+\mathbf{w}_k - \mathbf{x}_{k|k-1})(f_k(\mathbf{x}_{k-1})+ \bm{\Delta}_k^f + \mathbf{w}_k - \mathbf{x}_{k|k-1})^\mathrm{T}$.

Substistuting Eqs. \eqref{asdasd} - \eqref{delta_f}, \eqref{ex_pred_mis} - \eqref{dasdas} into Eq. \eqref{X_pppred} can be calculated as  
\begin{flalign}
    \mathbf{P}_{k|k-1} &= \int f_k(\mathbf{x}_{k-1})(f_k(\mathbf{x}_{k-1}))^\mathrm{T} P_{k-1}^x \mathrm{d}\mathbf{x}_{k-1} \nonumber \\
    & + \mathbf{x}_{k|k-1}(\mathbf{x}_{k|k-1})^\mathrm{T} + \mathbf{Q}_k + \mathbf{P}_k^f. \label{state_cov_prediction_temp}
\end{flalign}
Now, the first-order Taylor's expansion of \eqref{state_cov_prediction_temp} yields Eq. \eqref{state_cov_pred_final}.

The extension predicted probability density at time $k$ is computed as
\begin{align}
    & p(\mathbf{X}_k|\mathbf{z}_{1:k-1},\mathcal{D}) = \iint p(\mathbf{x}_k,\mathbf{X}_k,\mathbf{c}_k|\mathbf{z}_{1:k-1},\mathcal{D}) \mathrm{d}\mathbf{x}_k\mathrm{d}\mathbf{c}_k \nonumber \\
    &= \iiint \! \! \! \iint P_k^1 p(\mathbf{x}_{k-1}, \mathbf{X}_{k-1}, \mathbf{c}_{k-1}|\mathbf{Z}_{1:k-1}, \mathcal{D}) \mathrm{d}\mathbf{x}_{k-1} \nonumber \\
    & \quad \ \mathrm{d}\mathbf{X}_{k-1}\mathrm{d}\mathbf{c}_{k-1}\mathrm{d}\mathbf{x}_k\mathrm{d}\mathbf{c}_k. \label{cuhasfbyb}
\end{align}

Adopting the method used by \cite{feldmann2008tracking} to approximate $\mathbf{X}_{k-1}$ with $\mathbf{X}_{k|k-1}$, and utilizing the Cholesky decomposition technique, the transition process in Eq. \eqref{asdasd} can be equivalently expressed as
\begin{align}
    \mathbf{A}_k\mathbf{X}_k\mathbf{A}_k^\mathrm{T}+\bm{\Delta}_k^\phi &\approx (\mathbf{A}_k\mathbf{X}_{k|k-1}\mathbf{A}_k^\mathrm{T}+\bm{\Delta}_k^\phi)^{\frac{1}{2}}\mathbf{X}_{k|k-1}^{-\frac{1}{2}}\mathbf{X}_{k-1} \nonumber \\
    & \times \mathbf{X}_{k|k-1}^{-\frac{\mathrm{T}}{2}}(\mathbf{A}_k\mathbf{X}_{k|k-1}\mathbf{A}_k^\mathrm{T}+\bm{\Delta}_k^\phi)^{\frac{\mathrm{T}}{2}} \nonumber \\
    &= \mathbf{A}_k^*\mathbf{X}_k(\mathbf{A}_k^*)^\mathrm{T}. \label{asdasdwsdwsg}
\end{align}
\indent Substistuting Eqs. \eqref{asdasd}, \eqref{memoryasdasd}, \eqref{dasdas} and \eqref{asdasdwsdwsg} into Eq. \eqref{cuhasfbyb} can be calculated as
\begin{align}
    p(\mathbf{X}_k|\mathbf{z}_{1:k-1},\mathcal{D}) &= \! \! \iint \! \mathcal{GB}_d^{\mathrm{\uppercase\expandafter{\romannumeral2}}}(\mathbf{X}_k;\delta_k/2,\frac{v_{k-1|k-1}-d-1}{2}; \nonumber \\
    & \quad \quad \quad \! \! \mathbf{A}_k^*\mathbf{X}_k(\mathbf{A}_k^*)^\mathrm{T}, \mathbf{0}) \nonumber \\
    & \!\!\! \times p(\bm{\Delta}_k^\phi|\mathbf{c}_k,\mathcal{D})\mathcal{N}(\mathbf{c}_k;\hat{\mathbf{c}}_k,\mathbf{P}_k^c) \mathrm{d}\bm{\Delta}_k^\phi\mathrm{d}\mathbf{c}_k,
\end{align}
where $\mathcal{GB}_d^{\mathrm{\uppercase\expandafter{\romannumeral2}}}(\cdot)$ is the Generalized Beta Type \uppercase\expandafter{\romannumeral2} (GBII) distribution.

To obtain recursive estimation, we consider approximating this GBII distribution with an inverted Wishart distribution by
moment matching similarly as in \cite{koch2008bayesian}:
\begin{align}
    p(\mathbf{X}_k|\mathbf{z}_{1:k-1},\mathcal{D}) \approx \mathcal{IW}(\mathbf{X}_k;\mathbf{v}_{k|k-1}, \mathbf{X}_{k|k-1}),
\end{align}
with $\mathbf{v}_{k|k-1}$ and $\mathbf{X}_{k|k-1}$ are calculated as Eqs. \eqref{v_pred_final} and \eqref{extension_pred_final}, respectively.

The likelihood to get the measurement $\mathbf{Z}_k$ given both state and extension as well as the number of measurements in Eq. \eqref{likelihood}, can be factored as
\begin{gather}
    p(\mathbf{Z}_k|n_k, \bm{\Delta}_k^h,\mathbf{x}_k,\mathbf{X}_k) = \prod_{i=1}^{n_k} \mathcal{N}(\mathbf{z}_k^i; h_k(\mathbf{x}_k + \bm{\Delta}_k^h, \mathbf{B}_k \mathbf{X}_k \mathbf{B}_k^\mathrm{T}) \nonumber \\
    \!\!\!\!\!\!\!\! \propto \! \mathcal{N}(\widetilde{\mathbf{z}}_k;\!h_k(\mathbf{x}_k)\!+\!\bm{\Delta}_k^h,\!\frac{\mathbf{B}_k \mathbf{X}_{k}\mathbf{B}_k^\mathrm{T}}{n_k})\!\mathcal{W}(\widetilde{\mathbf{Z}}_k;\!n_k\!-\!1,\!\mathbf{B}_k\mathbf{X}_k\mathbf{B}_k^\mathrm{T}). \!\!\!\! \label{ll}
\end{gather}

For state prediction, the mean of its Gaussian measurement is computed as
\begin{align}
    \mathbf{z}_{k|k-1} & = \mathbb{E}_{\mathbf{x}_k|\widetilde{\mathbf{z}}_{1:k}, \mathcal{D}}[h_k(\mathbf{x}_k) + \bm{\Delta}_k^h] \nonumber \\
    & = \iiiint (h_k(\mathbf{x}_k) + \bm{\Delta}_k^h) P_k^3 \mathrm{d}\bm{\Delta}_k^h \mathrm{d}\mathbf{c}_k \mathrm{d}\mathbf{X}_k, \label{z_k_k-1}
\end{align}
\vspace{-0.5em}
with 
\begin{align}
    P_k^3 = p(\bm{\Delta}_k^h | \mathbf{x}_k, \mathbf{X}_k, \mathcal{D}) p(\mathbf{x}_k,\mathbf{X}_k,\mathbf{c}_k|\mathbf{Z}_{1:k-1},\mathcal{D}).
\end{align}

Substistuting Eq. \eqref{delta_h} into Eq. \eqref{z_k_k-1} then we have
\begin{gather}
    \mathbf{z}_{k|k-1} = \int h_k(\mathbf{x}_{k|k-1}) P_{k-1}^x \mathrm{d}\mathbf{x}_{k-1} + \hat{\bm{\Delta}}_k^h. \label{pred_meas_temp}
\end{gather}
whose  first-order Taylor's expansion leads to Eq. \eqref{state_pred_meas}.

We denote the residual between the measurement and the predicted measurement as $\bm{\sigma}_k^z = \widetilde{\mathbf{z}}_k - \mathbf{z}_{k|k-1}$, and the corresponding covariance of measurement error is given by
\begin{align}
    \mathbf{P}_{k|k-1}^{zz} &= \mathbb{E}_{\mathbf{x}_k|\widetilde{\mathbf{z}}_k, \mathcal{D}}[\bm{\sigma}_k^z(\bm{\sigma}_k^z)^\mathrm{T}].\label{zz}
\end{align}
Substistuting the Gaussian term in Eq. \eqref{ll} and Eq. \eqref{delta_h} into Eq. \eqref{zz} thus we have
Eq. \eqref{P_zz}.
The mutual covariance of the state prediction and the predicted measurement is given by
\begin{align}
    \mathbf{P}^{xz}_{k|k-1} &= \mathbb{E}_{\mathbf{x}_k|\widetilde{\mathbf{z}}_k, \mathcal{D}}[(\mathbf{x}_k - \mathbf{x}_{k|k-1})(\bm{\sigma}_k^z)^\top] 
\end{align}
Following the conclusions in \cite{wang2012gaussian}, we have the posterior state and covariance as Eqs. \eqref{state_update_final} and \eqref{state_cov_update_final}, respectively. The update of the extension follows the same derivation method as \cite{lan2012tracking}, leading to the Eqs. \eqref{v_update_final} and \eqref{extension_update_final}.

\bibliographystyle{IEEEtran}
\bibliography{IEEEabrv,EOTNet__Deep_Memory_Aided_Bayesian_Filter_for_Extended_Object_Tracking}

\end{document}